\documentclass[12pt]{article}
\usepackage{amsmath}
\makeatletter
\DeclareFontFamily{U}{mathx}{}
\DeclareFontShape{U}{mathx}{m}{n}{<-> mathx10}{}
\DeclareSymbolFont{mathx}{U}{mathx}{m}{n}
\DeclareMathAccent{\widecheck}{0}{mathx}{"71}
\makeatother

\usepackage{graphicx,float}
\usepackage{caption}
\usepackage{subcaption}
\usepackage{amsthm,bbm}
\usepackage{color}
\usepackage{mathrsfs}
\usepackage{mathtools}
\usepackage[authoryear]{natbib}
\usepackage[titletoc,title]{appendix}
\usepackage{amssymb,bm}
\usepackage{setspace}
\usepackage[flushleft]{threeparttable}
\usepackage{booktabs,etoolbox}
\usepackage{mathrsfs}
\usepackage{makecell}

\usepackage{natbib}
\setcitestyle{authoryear,open={(},close={)}}
\usepackage{hyperref}
\usepackage{dsfont}
\hypersetup{
  colorlinks = true, 
  urlcolor   = blue, 
  linkcolor  = blue, 
  citecolor  = blue  
}
\usepackage{comment}
\usepackage{multirow}
\usepackage{enumerate}

\usepackage{mathtools}

\DeclareMathOperator*{\argmin}{argmin}

\newtheorem{assumption}{Assumption}

\newtheorem{lemma}{Lemma}

\newtheorem{theorem}{Theorem}
\numberwithin{equation}{section}
\newtheorem{algorithm}{Algorithm}

\makeatletter
\renewenvironment{proof}[1][\proofname]{%
  \par\pushQED{\qed}%
  \normalfont\topsep6\p@\@plus6\p@\relax
  \trivlist
  \item[\hskip\labelsep \bfseries #1\@addpunct{.}]\ignorespaces
}{%
  \popQED\endtrivlist\@endpefalse
}
\makeatother

\usepackage{xr}
\makeatletter
\newcommand*{\addFileDependency}[1]{
\typeout{(#1)}
\@addtofilelist{#1}
\IfFileExists{#1}{}{\typeout{No file #1.}}
}\makeatother

\title{Quantile Treatment Effects in High Dimensional Panel Data
}
\author{Yihong Xu\footnote{Department of Economics, Texas A\&M University, US. Email: kaiserxu\_96@tamu.edu.}~ and Li Zheng\footnote{Corresponding author. Institute for Economic and Social Research, Jinan University, China. Email: lizheng.gz@hotmail.com.}}
\date{\today}

\begin{document}

\maketitle
\begin{abstract}
    We introduce novel estimators for quantile causal effects with high-dimensional panel data (large $N$ and $T$), where only one or a few units are affected by the intervention or policy. Our method extends the generalized synthetic control method \citep{xu_2017} from average treatment effects on the treated to quantile treatment effects on the treated, allowing the underlying factor structure to change across the quantile of the interested outcome distribution. Our method involves estimating the quantile-dependent factors using the control group, followed by a quantile regression to estimate the quantile treatment effect using the treated units. We establish the asymptotic properties of our estimators and propose a bootstrap procedure for statistical inference, supported by simulation studies. An empirical application of the 2008 China Stimulus Program is provided. 

    ~

\end{abstract}

\newpage

\section{Introduction}
\label{sec:intro}
Over the past two decades, Synthetic Control Method (SCM) \citep{abadie2003economic,abadie2010synthetic} has been extensively applied in empirical research in economics and social sciences. It estimates the average treatment effects on the treated (ATT) by constructing a counterfactual of a single or small group of treated units using the control group. Due to its interpretability, transparency, and plausible identification assumptions, SCM has gained popularity among practitioners, and is considered “arguably the most important innovation in the policy evaluation literature in the last 15 years” \citep{athey2017state}. In particular, \cite{xu_2017} proposes the generalized synthetic control method (GSCM) adapting SCM to a high dimensional panel data setting (large $N$ and $T$) with numerous control units. 
This approach naturally incorporates a factor model to manage the abundance of control units, thereby attracting significant interest for estimating ATT in high dimensional panel data.

Beyond the mean effect, researchers might be concerned with the distributional changes caused by certain policies or interventions. 
For example, \cite{adrian2019vulnerable} argue that traditional economic forecasting, which focuses primarily on the conditional mean of GDP growth, tends to obscure the importance of downside risks. They propose assessing the vulnerability of economic growth by examining the full conditional distribution of GDP, with a particular emphasis on its lower tails. Such distinct behaviors across different parts of the distribution of GDP growth raise a natural question: How will a macroeconomic policy affect different quantiles of the GDP growth? In our empirical application, where we study the impact of 2008 Chinese Economic Stimulus Program on GDP growth, we try to answer the following question: How will this policy alter GDP growth during economic downturns (lower quantiles) or booms (upper quantiles)?

In this paper, we consider the quantile treatment effects on the treated (QTT) in the SCM context with high dimensional panel data, where we have a single or a few treated units and a large number of control units. QTT delivers the shifts in outcome distributions caused by a policy at different quantiles, allowing one to fully characterize the impact of the intervention. 
Our estimators take a similar spirit to \cite{xu_2017}'s GSCM, which posits a factor model \citep{bai2003inferential,bai2009panel} to capture the unobserved time-variant heterogeneities on the panel data. The GSCM first estimates the unknown factors using the control group, then estimates the factor loadings for the treated units, and finally imputes the counterfactual outcome for the treated units to calculate the ATT. 
To study the QTT, we consider a quantile factor model \citep[QFM]{chen2021quantile}, which allows the number and values of the factors as well as its loadings to vary across different quantiles. This setting is motivated by the increasing evidence of comovements of some economics and finance variables \citep{adrian2019vulnerable,de2019dynamic} and admits much more flexibility in the model.
Our QTT estimation proceeds in two stages. First, we estimate the unobserved quantile-dependent factors using the control units, with a data-driven method to determine the number of factors, following \cite{chen2021quantile}. Next, we take the estimated factors into quantile regressions on the treated units to deliver QTT estimates. Our proposal complements the flourishing SCM literature by providing methodologies for evaluating distributional causal effects.

We consider two QTT estimators differing at their estimation of common factors in the first stage, where the first one relies on a nonsmooth objective function (the check function) and the second one is based on a smoothed alternative. We refer to them as NQTT (Nonsmooth QTT) and SQTT (Smoothed QTT), respectively. At the cost of more computational burden, SQTT achieves a faster convergence rate of the estimated factors, lending us the inference theory for the final QTT estimator. We prove the consistency for NQTT and the asymptotic normality for SQTT. For SQTT, we provide a blockwise bootstrap method \citep{horowitz2019bootstrap} to construct confidence intervals, since the asymptotic variance might not be accurately estimated, as is well addressed in quantile regressions \citep{koenker2017handbook}. 
We further establish the consistency of our bootstrap procedure, offering a theoretical justification for its validity. Monte-Carlo simulations are conducted to demonstrate a good finite sample performance of our method. To illustrate the implementation of our proposal, we apply it to analyze the macroeconomic effects of the 2008 Chinese Economic Stimulus Program \citep{ouyang2015treatment}. Our QTT estimation supports that the stimulus package has a significant effect on GDP growth for quantiles lower than 50\%. This finding reinforces the widely circulated perception that China's fiscal policies in 2008 served as a safety net, preserving the bottom line of economic growth and ensuring that the Chinese economy could maintain high-speed growth even amidst unfavorable global economic conditions. 
Finally, we discuss several possible extensions for our method, including time varying QTT, multiple treated units, and the use of covariates.

Despite its importance, the literature on quantile treatment effects within the SCM framework is relatively sparse. 
\cite{chen2020distributional} extends the SCM \citep{abadie2003economic} to evaluate the distributional effects of policy interventions in the possible presence of poor matching, but the ``distributional" information comes from subunits (or individuals) within an aggregate-level unit. That is, researchers must observe $\{y_{i\ell t}\}_{\ell=1}^L$ to identify the distribution of unit $i$ at time $t$. \cite{gunsilius2023distributional} imposes a similar data requirement and constructed counterfactual distribution for the treated unit by taking a weighted average of the untreated $y_{it}$'s distributions, while \cite{chen2020distributional} focuses on a given quantile. This line of methodologies, while innovative, may not always be applicable in empirical settings where we only observe the aggregate level outcome $y_{it}$, such as in cases involving macroeconomic \citep[GDP growth]{adrian2019vulnerable}, or financial \citep[stock return]{ando2020quantile} variables. 
\cite{cai2022estimating} also consider quantile treatment effects estimation in panel data building on \cite{hsiao2012panel}. They propose to construct the counterfactual distribution with the pre-treatment data by first nonparametrically estimating the conditional cumulative distribution function (CDF), and then transforming it into unconditional CDF to obtain the estimated quantile function.
However, this method may falter when the total number of covariates and control units is large due to the curse of dimensionality of the nonparametric estimation. When facing a high dimensional panel data, they provide an alternative estimation scheme that relies on penalized quantile regressions, whereas its theoretical properties are underexplored.

Compared to existing works, our method has the following advantages. 
Firstly, it can be directly extended to the multiple treated units cases. \cite{abadie2021using} highlighted the challenge that classical matching methods with simplex conditions may provide multiple (even infinitely many) solutions. This would lead to a practical issue of choosing proper weights, when the multiple treated units have quite different characteristics. 
Secondly, our method effectively handles the high-dimensionality in panel data, as the factor models are inherently designed for dimension reduction. Therefore, we are able to capture the common trends from a large number of control units and use them to identify the underlying causal effects on the treated unit. 
Finally, compared to the standard factor models as in \cite{bai2009panel} and \cite{xu_2017}, QFM captures hidden factors that may shift characteristics (moments or quantiles) of the outcome distribution beyond its mean, permitting the factors, factor loadings, and number of factors to vary across the quantile levels of the outcome distribution.

The remainder of the paper is organized as follows. Section \ref{sec:framework} describes the model, outlines the estimation procedure, and proposes the blockwise bootstrap method to construct confidence intervals. Section \ref{sec:asymptotic} establishes the asymptotic theory for the proposed estimators. Section \ref{sec:simulation} shows the results of the Monte Carlo simulation. Section \ref{sec:empirical} illustrates the empirical implementation of our method by evaluating the policy effect of the 2008 Chinese Economic Stimulus Program. Section \ref{sec:extension} explores potential extensions of our estimators. Section \ref{sec:conclusion} concludes. The proofs as well as additional simulation and empirical results are presented in the Appendix.

\section{Framework}
\label{sec:framework}

\subsection{Model}
We posit the potential outcome framework in a panel data environment. Suppose $d_{it}\in\{0,1\}$ is a binary treatment indicating whether unit $i=1,\cdots,N,N+1$ is treated at time $t=1,\cdots,T$. Let $y^1_{it}$ and $y^0_{it}$ denote the potential outcomes under and in the absence of treatment, respectively. Following the convention in the SCM literature, we consider the case that only the first unit $i=1$ receives the treatment which occurs at time $T_0+1$, and continues till the last time period $T$. Denote $T_1=T-T_0$ as the number of treated periods. Resembling the Stable Unit Treatment Value Assumption (SUTVA) \citep{rubin1980randomization}, we suppose there is no ``anticipation effect" so that the treatment does not affect the treated unit in the pretreatment period, and there is no ``interference effect" so that the control units are not influenced by the treatment \citep{abadie2010synthetic}. Therefore, the observed outcome is given by $y_{it} = d_{it}y^1_{it} + (1-d_{it})y^0_{it}$.

Generally, researchers are interested in ATT, i.e., $\mathbb{E}[y_{1t}^1-y_{1t}^0]$ for $t>T_0$, and numerous ATT estimators and their properties have been developed in the SCM literature, e.g., \cite{abadie2010synthetic}, \cite{hsiao2012panel}, \cite{xu_2017}, and \cite{carvalho2018arco}, to name a few. If, however, the distributions of potential outcomes poorly concentrate around the mean, the distributional change of the outcomes, instead of ATT, would better represent the treatment effects. Next, we will present our model and the QTT estimand.

We consider that the correlations among cross-sectional units are due to some common factors that drive all cross-sectional units, whose impacts on each cross-sectional unit may be different, and allow that hidden factors may shift characteristics (moments or quantiles) of the distribution of outcome variable. Moreover, the common factors, factor loadings, as well as the number of factors could vary across distributional characteristics. Therefore, we introduce the following QFM \citep{chen2021quantile} at some quantile $\tau \in (0,1)$ :
\begin{equation}
\label{qfs_linear}
\begin{split}
    Q_{\tau}(y^0_{it}|f_t(\tau)) & =\lambda_i'(\tau)f_t(\tau), \\
    Q_{\tau}(y^1_{it}|f_t(\tau)) & =\delta(\tau) + \lambda_i'(\tau)f_t(\tau), 
\end{split}
\end{equation}
where $Q_{\tau}(y|x)$ denotes the $\tau$-th quantile of $y$ conditional on $x$. Here $f_t(\tau)$ is an $r(\tau) \times 1$ vectors of common factors, $\lambda_i(\tau)$ is an $r(\tau) \times 1$ vector of factor loadings, with $r(\tau) \ll N$. The notation of $f_t(\tau)$, $\lambda_i(\tau)$ and $r(\tau)$ indicates that they are all allowed to be quantile-dependent. Our causal parameter of interest QTT is $\delta(\tau)$ as defined in (\ref{qfs_linear}), as in \cite{firpo2007efficient}, \cite{chernozhukov2005iv}, etc. If $y_{it}$ is GDP growth and $d_{it}$ is the economic stimulus program, then QTT $\delta(\tau)$ represents the difference of the $\tau$-th quantile of GDP growth distribution between receiving and not receiving the program. For a small $\tau$, $\delta(\tau)$ reflects the policy impact under economic downturns. Notice that the model would turn to \cite{xu_2017}'s GSCM setting when one replaces the quantile operators with expectation operators.
The above QFM can be written as: 
\begin{align} \label{qfs_linear2}
    y_{it} =\delta(\tau)d_{it} + \lambda_i'(\tau)f_t(\tau) + \epsilon_{it}(\tau), 
\end{align}
where $\epsilon_{it}(\tau)$ has zero $\tau$-th quantile conditional on $f_t(\tau)$ , i.e. $Q_{\tau}(\epsilon_{it}|f_t(\tau))=0$.

Consider a DGP used in our simulation as an illustrative example:
\begin{align}
\label{eqn:example}
   y^0_{it} = \lambda_{1i}f_{1t} + \lambda_{2i}f_{2t}  + \lambda_{3i}f_{3t} u_{it}, \,\, y_{1t}^1=y_{1t}^0+\alpha_t,
\end{align}
where $\alpha_{t} = u_{1t} + 0.5$ and $u_{it}\stackrel{\text{i.i.d}}{\sim}\mathcal{N}(0,1)$. The mean of $y^0_{it}$ is determined by $f_{1t}$ and $f_{2t}$, while the quantile of $y^0_{it}$ is additionally determined by $f_{3t}$. We may fit this location-scale shift model for $y_{it}^0$ into QFM by writing $\lambda_i(\tau)=(\lambda_{1i},\lambda_{2i})'$, $f_t(\tau) = (f_{1t},f_{2t})'$ if $\tau= 0.5$, and $\lambda_i(\tau)=(\lambda_{1i},\lambda_{2i},\lambda_{3i} \Phi^{-1}(\tau))'$, $f_t(\tau) = (f_{1t},f_{2t},f_{3t})'$ if $\tau\neq 0.5$. The standard factor model would fail to capture $f_{3t}$ at $\tau \neq 0.5$, while QFM is capable of describing such a quantile-dependent feature. In this case, $\epsilon_{it}(\tau) = y^0_{it} - Q(y^0_{it} \mid f_t)=\lambda_{3i} f_{3t} (u_{it}-Q_\tau(u_{it}|f_t))$.

In the above example, we see that the true QTT is $\delta(\tau) = 0.5 + \Phi^{-1}(\tau)$, where $\Phi$ is the CDF function of the standard normal distribution. Although the QTT varies over $\tau$ and reflects the distributional changes due to the treatment, it remains constant for a given quantile. Different from \cite{chen2020distributional} and \cite{gunsilius2023distributional} who could utilize subunit level data $\{y_{1\ell t}\}_{\ell=1}^L$ to obtain the quantile functions $Q_\tau(y_{1t})$ and identify QTT $\delta_t$ for each $t$, we only observe the aggregate outcome $y_{1t}$, so it is only possible to identify a finite-dimensional parameter for the distribution of treatment effect.\footnote{In this example, the quantile of treatment effect $\alpha_t$ equals to our defined QTT. See \citet[Theorem 24.5]{hansen2022econometrics} for further discussions on the definitions of quantile treatment effects.} In this sense, we share a similar spirit with \cite{cai2022estimating}. However, our quantile regression framework in the second stage could be extended to allow QTT to be functions of time $t$ or factors $f_t$, which enables a parametric dependence of QTT on time. We discuss this extension in Section \ref{sec:relax-linear}.

\subsection{Estimation}

We propose our QTT estimator in this subsection. We first estimate the common factors in QFM using the control units following \cite{chen2021quantile}, with a data-driven method to determine the factor numbers. Next, we estimate QTT by quantile regressions based on the estimated factors using the treated unit.

To simplify the notations, we suppress hereafter the dependence of $\delta(\tau)$, $f_t(\tau)$, $\lambda_i(\tau)$, $r(\tau)$ and $\epsilon_{it}(\tau)$ on $\tau$. Denote the true values of $\{f_t\}$, $\{\lambda_i\}$, and $\delta$ as $\{f_{0t}\}$, $\{\lambda_{0i}\}$, and $\delta_0$, respectively. Let $\Delta\subset \mathbb{R}$ be the parameter space for $\delta$. Following \cite{chen2021quantile}, we treat $f_{0t}$ and $\lambda_{0i}$ as parameters to be estimated. Let $\theta_0=(\theta_{0\lambda}',\theta_{0f}')'=(\lambda_{01}',\lambda_{02}',...,\lambda_{0N}',f_{01}',...,f_{0T}')'$, and $\theta=(\theta_{\lambda}',\theta_{f}')'=(\lambda_1',\lambda_2',...,\lambda_N',f_1',...,f_T')'$. 
Following the convention in the factor model literature, we impose normalization conditions for the identification of factors and loadings for any positive integer $T$  and $N$, i.e.,
\begin{align}
    \label{norm1}
    &\frac{1}{T}\sum_{t=1}^Tf_tf_t'=\mathbb{I}_r,\,\,\,
    \frac{1}{N}\sum_{i=2}^{N+1}\lambda_i\lambda_i' \mbox{ is diagonal with non-increasing diagonal elements.}
\end{align}
Let $M=(N+T)r$, $\mathcal{A},\mathcal{F}\subset\mathbb{R}^r$, and define
\begin{align}
\Theta^r=\{\theta\in\mathbb{R}^M:\lambda_i\in\mathcal{A},f_t\in\mathcal{F} \mbox{ for all }i,t,\{\lambda_i\} \mbox{ and }\{f_t\} \mbox{ satisfy \eqref{norm1}}\}. 
\end{align}

\subsubsection{Estimating Common Factors}

\cite{chen2021quantile} propose two estimators for the common factors in the QFM. The first method considers the minimization of the following objective function.
\begin{align}\label{obj_fuc_1}
   \mathbb{M}_{NT}(\theta)=\frac{1}{NT} \sum_{i=2}^{N+1}\sum_{t=1}^T \rho_{\tau}(y_{it}-\lambda_i'f_t),  
\end{align}
where $\rho_{\tau}(a)=a(\tau-\mathbf{1}(a \leq 0))$ is the check function.
We estimate $\theta_0$ by $\widehat{\theta}= \underset{\theta\in\Theta^r}{\text{argmin }} \mathbb{M}_{NT}(\theta)$ where the common factors $f_{0t}$ are a part of $\theta_0$. Since $\mathbb{M}_{NT}(\theta)$ is not convex in $\theta$, \cite{chen2021quantile} introduce the Iterative Quantile Regression (IQR) algorithm. Let $F=(f_1,...,f_T)'$, $\Lambda=(\lambda_1,\lambda_2,...,\lambda_N)'$ and
\begin{align}\label{obj_fuc_2}
   \mathbb{M}_{i,T}(\lambda,F)=\frac{1}{T} \sum_{t=1}^T \rho_{\tau}(y_{it}-\lambda_i'f_t), \hspace{0.3cm}  \mathbb{M}_{N,t}(\Lambda,f)=\frac{1}{N} \sum_{i=2}^{N+1} \rho_{\tau}(y_{it}-\lambda_i'f_t).
\end{align} 
Based on the fact that $\mathbb{M}_{i,T}(\lambda,F)$ is convex in $\lambda$ for each $i$, and $\mathbb{M}_{N,t}(\Lambda,f)$ is convex in  $f$ for each $t$, an iterative procedure of optimizations for $\mathbb{M}_{i,T}(\lambda,F)$ and $\mathbb{M}_{N,t}(\Lambda,f)$ could lead to the optimization of $\mathbb{M}_{NT}(\theta)$.  Algorithm \ref{algo-IQR} illustrates the IQR algorithm for a given $\tau$ and $r$.

\begin{algorithm}[Iterative Quantile Regression (IQR)] 
\label{algo-IQR}
\hfill\par
\begin{enumerate}[Step 1:]
    \item Select a random starting factor: $F^{(0)}$; 
    \item Given $F^{(l-1)}$, solve $\lambda_i^{(l-1)}= \underset{\lambda}{\argmin}\, \mathbb{M}_{i,T}(\lambda,F^{(l-1)})$ for $i=2,...,N+1$; Given $\Lambda^{(l-1)}$, solve $f_t^{(l)}= \underset{f}{\argmin}\, \mathbb{M}_{N,t}(\Lambda^{(l-1)},f)$ for $t=1,...,T$;
    \item For $l=1,...,L$, iterate Step 2 until $\mathbb{M}_{NT}(\theta^{(l-1)})$ is close enough to $\mathbb{M}_{NT}(\theta^{(l)})$;
    \item Normalize $F^{(L)}$ and $\Lambda^{(L)}$ according to \eqref{norm1}, and denote them as $\widehat{F}$ and $\widehat{\Lambda}$.
\end{enumerate}
\end{algorithm}

While the IQR algorithm is computationally efficient and convenient, its convergence rate for the estimated factors $\widehat{f}_t$ is only sufficient for the consistency, but not for the asymptotic normality of the QTT estimator in the second stage. In particular, as shown in the proof of our Theorem \ref{thm2}, the asymptotic normality result requires that
\begin{align}
\label{eqn:rate-fhat}
    \frac{1}{\sqrt{T}}\sum_{t=1}^T (\widehat{f}_t-\widehat{S}f_{0t})'c_t = o_p(1),
\end{align}
where $c_t$ is a bounded sequence, and $\widehat{S}=\operatorname{sgn}(\widehat{F}'F/T)$. However, \cite{chen2021quantile} shows that $T^{-1/2}\sum_{t=1}^T\|\widehat{f}_t-\widehat{S}f_{0t}\|^2=O_p(L_{NT}^{-1})$ where $L_{NT}=\min\{\sqrt{N},\sqrt{T}\}$, so that one can only show
\begin{align}
    \frac{1}{\sqrt{T}}\sum_{t=1}^T (\widehat{f}_t-\widehat{S}f_{0t})'c_t \leq \bigg(\frac{1}{T}\sum_{t=1}^T\|\widehat{f}_t-\widehat{S}f_{0t}\|^2\bigg)^{1/2}\bigg(\sum_{t=1}^Tc_t\bigg)^{1/2} = \sqrt{T}O_p(L_{NT}^{-1}),
\end{align}
which is at best $O_p(1)$. The underlying reason is the non-smoothness of the check function in the objective function \eqref{obj_fuc_1}. \cite{bai2008extremum} also considers M-estimation where the regressors are estimated from the standard factor models \citep{bai2003inferential}, and establish asymptotic normality for the regression coefficients. However, they manage to achieve the rate in \eqref{eqn:rate-fhat} using the stochastic expansion of $\widehat{f}_t-\widehat{S}f_{0t}$ in standard factor models \citep{bai2002determining}, and its extension to QFM based on IQR appears not trivial \citet[p.886-887]{chen2021quantile}.

In view of that, we consider a second method for estimating $f_{0t}$ proposed in \cite{chen2021quantile} which admits asymptotic normality for the QTT estimator. It replaces the non-smooth objective function \eqref{obj_fuc_1} by a smoothed alternative:
\begin{align}
     \widetilde{\mathbb{M}}(\theta)=\frac{1}{NT} \sum_{i=2}^{N+1}\sum_{t=1}^T \left[ \tau - K\bigg(\frac{y_{it}-\lambda_i'f_t}{h}\bigg)\right](y_{it}-\lambda_i'f_t),
\end{align}
where $K(z)=1-\int_{-1}^z k(z) dz$, $ k(z)$ is a continuous kernel function with support $[-1,1]$, and $h$ is a bandwidth parameter. Define $ \widetilde{\mathbb{M}}_{i,T}$ and $ \widetilde{\mathbb{M}}_{N,t}$ as the smoothed version of \eqref{obj_fuc_2}:
\begin{equation}
\begin{split}
   & \widetilde{\mathbb{M}}_{i,T}(\lambda,F)=\frac{1}{T} \sum_{t=1}^T \left[ \tau - K\bigg(\frac{y_{it}-\lambda_i'f_t}{h}\bigg)\right](y_{it}-\lambda_i'f_t),\\
   & \widetilde{\mathbb{M}}_{N,t}(\Lambda,f)=\frac{1}{N} \sum_{i=2}^{N+1} \left[ \tau - K\bigg(\frac{y_{it}-\lambda_i'f_t}{h}\bigg)\right](y_{it}-\lambda_i'f_t),
\end{split} 
\end{equation}
and the estimation procedure is just modifying Algorithm \ref{algo-IQR} by using the smoothed objective functions.
\begin{algorithm}[Iterative Smoothed Quantile Regression (ISQR)] 
\label{algo-ISQR}
\hfill\par
\begin{enumerate}[Step 1:]
    \item Select a random starting factor: $\widetilde{F}^{(0)}$; 
    \item Given $\widetilde{F}^{(l-1)}$, solve $\widetilde{\lambda}_i^{(l-1)}= \underset{\lambda}{\argmin}\, \widetilde{\mathbb{M}}_{i,T}(\lambda,\widetilde{F}^{(l-1)})$ for $i=2,...,N+1$; Given $\widetilde{\Lambda}^{(l-1)}$, solve $\widetilde{f}_t^{(l)}= \underset{f}{\argmin}\, \widetilde{\mathbb{M}}_{N,t}(\widetilde{\Lambda}^{(l-1)},f)$ for $t=1,...,T$;
    \item For $l=1,...,L$, iterate Step 2 until $\widetilde{\mathbb{M}}_{NT}(\widetilde{\theta}^{(l-1)})$ is close enough to $\widetilde{\mathbb{M}}_{NT}(\widetilde{\theta}^{(l)})$;
    \item Normalize $\widetilde{F}^{(L)}$ and $\widetilde{\Lambda}^{(L)}$ according to \eqref{norm1}, and denote them as $\widetilde{F}$ and $\widetilde{\Lambda}$.
\end{enumerate}
\end{algorithm}
In Lemma \ref{lemma1} in the Appendix it can be seen that the estimated factors $\widetilde{f}_t$ possess the desired convergence rate in \eqref{eqn:rate-fhat}. In the next section, we show that QTT estimator based on $\widetilde{f}_t$ is asymptotically normal which facilitates valid statistical inference.

\subsubsection{Estimating the Number of Factors}

While in the above algorithms, $r$ is supposed to be known, the researcher needs to first determine it and usually prefer a data-driven method to do so. Let $r_0$ denote the true factor number, and $P_{NT}$ to be a sequence goes to 0 as $N,T\rightarrow \infty$. 
\cite{chen2021quantile} use Algorithm \ref{algo-MSRM} to choose a proper factor number by rank minimization, and show that it produces a consistent estimator for $r_0$.

\begin{algorithm}[Estimating the Number of Factors] 
\label{algo-MSRM}
\hfill\par
\begin{enumerate}[Step 1:]
    \item Choose a positive integer $k$ that is big enough ($k > r_0$); 
    \item Get $\widehat{\Lambda}^k=(\widehat{\lambda}^k_1,\widehat{\lambda}^k_2,...,\widehat{\lambda}^k_N)'$ using Algorithm \ref{algo-IQR}, where the upper script means the estimation is based on $r = k$; Write $(\widehat{\Lambda}^k)'\widehat{\Lambda}^k /N = \operatorname{diag}(\widehat{\sigma}_{N,1}^k,...,\widehat{\sigma}_{N,k}^k)$;
    \item The estimator of the number of factors is $\widehat{r}_{RM} = \sum_{j=1}^k \mathbf{1}(\widehat{\sigma}_{N,j}^k \geq P_{NT})$.
\end{enumerate}
\end{algorithm}

The intuition of the rank minimization is, if we select a number of factors $k$ larger than the true $r_0$, as $N$ and $T$ grows, only the largest $r_0$ diagonal elements of $(\widehat{\Lambda}^k)'\widehat{\Lambda}^k /N$ would stay positive, while the rest $k-r_0$ diagonal elements would converge to zero. The efficiency of the rank minimization comes from the fact that one single proper value of $k$ leads to a consistent estimator $\widehat{r}_{RM}(\tau)$, compared to other information criterion techniques. Though we fail to observe the $r_0$ in a real world, we always assume the true factor numbers, under all possible quantiles, should be ``small" enough. In practice, \cite{chen2021quantile} suggest $P_{NT} = \widehat{\sigma}_{N,1}^k \cdot L_{NT}^{-2/3}$, where $ L_{NT} = \min\{\sqrt{N},\sqrt{T}\}$, and  $k \geq 8$ is recommended.

\subsubsection{Estimating the Quantile Treatment Effects}
Now we are ready to move to the estimation of the QTT. From Algorithm \ref{algo-IQR} and \ref{algo-ISQR} we have the estimated common factors ${\widecheck{f}_t}$, where we use ${\widecheck{f}_t}$ to denote $\widehat{f}_t$ or $\widetilde{f}_t$. To find the QTT, we simply conduct a quantile regression of the outcome on the estimated factors as well as the treatment, using the treated unit data. For a given quantile $\tau$, our QTT estimators is summarized in Algorithm \ref{algo-QTE}.

\begin{algorithm}[Estimating Quantile Treatment of Treated (QTT)]
\label{algo-QTE} 
\hfill\par
\begin{enumerate}[Step 1:]
    \item Select the number of factors $\widehat{r}_{RM}$ with Algorithm \ref{algo-MSRM}, and then estimate the quantile-dependent factors $\widecheck{f}_{t}$ using Algorithm \ref{algo-IQR} or \ref{algo-ISQR} with $\widehat{r}_{RM}$;
    \item With $\widecheck{f}_{t}$, estimate the quantile treatment effects $\widecheck{\delta}$ with the quantile regression:
    \begin{align} \label{QTEstep2}
        (\widecheck{\lambda}_{1},\widecheck{\delta}) = \underset{(\lambda_1,\delta)\in\mathcal{A}\times\Delta}{\operatorname{argmin}} \sum_{t=1}^T \rho_{\tau}(y_{1t}-\lambda_1'\widecheck{f}_{t}-\delta d_{1t}).
    \end{align}

\end{enumerate}
\end{algorithm}

\subsection{Inference}
\label{sec:inference}
Though the asymptotic normality of the QTT estimator $\widetilde{\delta}(\tau)$ based on ISQR is established in the next section, a direct estimation of the asymptotic variance could be unstable. This is because the asymptotic variance involves the conditional density function of the idiosyncratic errors (see Theorem \ref{thm2} below), whose nonparametric estimation might be inaccurate in finite samples \citep{koenker2017handbook}. Following the quantile regression literature, we use bootstrap method for inference. Since the common factors $f_{0t}$ might be time dependent, we apply the blockwise bootstrap \citep{kunsch1989jackknife} which is developed for inferences with time series data and has been adopted in quantile regressions \citep{gregory2018smooth}. Compared to the conventional pairwise bootstrap, the blockwise bootstrap respects the sequential order of the data by putting a certain number of sequential observations into one block, and taking random draws among the blocks. Our bootstrap inference procedure is summarized in Algorithm \ref{algo-blockboot}.

\begin{algorithm}[Blockwise Bootstrap] 
\label{algo-blockboot} 
\hfill\par
\begin{enumerate}[Step 1:]
    \item Define the the block size for pre-treatment data as $ \mathscr{B}_0= \lfloor \sqrt[3]{T_0}  \rfloor$ , and construct blocks as $\mathbb{B}_t=\{(y_{1t},\widetilde{f}_{t}'), \dots, (y_{1,t+\mathscr{B}_0-1},\widetilde{f}_{t+\mathscr{B}_0-1}') \}$. The total block set is $\mathbb{B}=\{\mathbb{B}_1, \dots, \mathbb{B}_{T_0-\mathscr{B}_0+1} \}$. Draw $\mathscr{L}_0 = \lfloor T_0/\mathscr{B}_0  \rfloor $ samples with replacement from $\mathbb{B}$, and denote the bootstrap sample by $V^\ast_0$;

    \item Conduct a similar bootstrap construction with $\mathscr{B}_1= \lfloor \sqrt[3]{T_1} \rfloor$ and $\mathscr{L}_1 = \lfloor T_1/\mathscr{B}_1  \rfloor $ for the post-treatment data. Denote the bootstrap sample by $V^\ast_1$;

    \item Run the quantile regression for a given $\tau$ in the complete bootstrap sample
    $(V^\ast_0,V^\ast_1)$, and estimate the QTT via \eqref{QTEstep2}.
    \item Repeat the above steps for $B$ times, and obtain a bootstrap sample for the QTT estimators: $(\widetilde{\delta}^{\ast(1)},\widetilde{\delta}^{\ast(2)}, ..., \widetilde{\delta}^{\ast(B)})$. Calculate the bootstrap sample standard deviation $\widetilde{\sigma}^\ast$, and the 95\% confidence interval can be constructed as $[\widetilde{\delta}-1.96\widetilde{\sigma}^\ast, \widetilde{\delta}+1.96\widetilde{\sigma}^\ast]$.
\end{enumerate}
\end{algorithm}

In the next section, we show that this blockwise bootstrap is consistent, in the sense that as $T\to\infty$, the bootstrap distribution becomes close to the asymptotic distribution of the QTT estimator, so that the confidence interval constructed from the bootstrap sample standard deviation is valid.

\section{Asymptotic Theory}
\label{sec:asymptotic}

In this section, we provide asymptotic properties for the proposed QTT estimators. All proofs are presented in the Appendix. We first elaborate on the required assumptions. Denote $w_{0t}=(f_{0t}',d_{1t})'$. 

\begin{assumption}[Strong Factor and Compactness]
\label{asmp-factor}
$\mathcal{A}$, $\mathcal{F}$, and $\Delta$ are compact sets and $\theta_0\in\Theta^r$. Moreover, $N^{-1}\sum_{i=2}^{N+1}\lambda_{0i}\lambda_{0i}'=\operatorname{diag}(\sigma_{N1},...,\sigma_{Nr_0})$ with $\sigma_{N1}\geq\cdots\geq\sigma_{Nr_0}$, and $\sigma_{Nj}\to\sigma_j$ as $N\to\infty$ for $j=1,...,r_0$ with $\infty>\sigma_1>\cdots>\sigma_{r_0}>0$.
\end{assumption}

\begin{assumption}[Conditional Density of Errors]
\label{asmp-density}
Denote the conditional density function of $\epsilon_{it}$ given $f_{0t}$ as $h_{it}(\epsilon)$. 
\begin{enumerate}[(i)]
    \item $h_{it}(\epsilon)$ is continuous in $\epsilon$ and satisfies that: for any compact set $C\subset \mathbb{R}$ and any $\epsilon\in C$, there exists a positive constant $\underbar{h}$ (depending on $C$) such that $h_{it}(\epsilon)\geq \underbar{h}$ for all $i=2,...,N+1$ and $t=1,...,T$, and for all $f_{0t}\in\mathcal{F}$.
    \item $h_{1t}(\epsilon)$ is continuous in $\epsilon$ and satisfies $h_{1t}(\epsilon)\leq \bar{h}<\infty$ for all $t=1,...,T$ and $f_{0t}\in\mathcal{F}$.
\end{enumerate}
\end{assumption}

\begin{assumption}[Mutually Independent Errors]
\label{asmp-error}
Given $\{f_{0t},1\leq t\leq T\},\{\epsilon_{it},2\leq i \leq N+1,1\leq t\leq T\}$ are independent across $i$ and $t$.
\end{assumption}

\begin{assumption}[Weak Dependency]
\label{asmp-weakdep}
$\{\epsilon_{1t}\}_{t=1}^T$ is a strictly stationary and $\alpha$-mixing sequence with mixing coefficient $\alpha(\ell)$ such that $\sum_{\ell=1}^{\infty}\alpha(\ell)^{1-\kappa}<\infty$ for some $\kappa>0$.
\end{assumption}

\begin{assumption}[Identification]
\label{asmp-iden}
\begin{enumerate}[(i)]
    \item $\{\epsilon_{1t}|f_{0t},d_{1t}=1\}\stackrel{d}{\sim}\{\epsilon_{1t}|f_{0t},d_{1t}=0\}$ for all $f_{0t}\in\mathcal{F}$.
    \item $\Gamma=\lim_{T\to\infty}T^{-1}\sum_{t=1}^Th_{1t}(0)w_{0t}w_{0t}'$ is positive definite.
    \item $\eta=\lim_{T\to\infty}T_0/T\in(0,1)$.
\end{enumerate}
\end{assumption}

Assumption \ref{asmp-factor} is the strong factor assumption which is satisfied when each factor has a non-trivial contribution to the outcome, and we can order the factors by the distinct $\sigma_j$. This is necessary for the true number of factors to be $r_0$, which is standard in the factor model literature \citep{bai2003inferential}. 
Assumption \ref{asmp-density} regulates the density functions of the idiosyncratic errors conditional on true factors. In particular, Assumption \ref{asmp-density}(i) requires that the conditional densities of errors are all positive for the control units, but does not require their moments to exist. This allows for heavy-tailed idiosyncratic errors such as standard Cauchy or some Pareto distributions. In the simulation study in Appendix, we show that our method is robust to a $t(2)$ distributed error. Assumption \ref{asmp-density}(ii) states that for the treated unit, the conditional densities are uniformly upper bounded, which is satisfied by many commonly used distributions. 
Assumption \ref{asmp-error} restricts the idiosyncratic errors of the control units to be mutually independent given the common factors. However, cross-sectional and temporal correlations among $\epsilon_{it}$ driven by common factors are still admitted. Moreover, we allow for cross-sectional and serial heteroscedasticity among $\epsilon_{it}$, making it a more general framework.
Assumption \ref{asmp-factor}-\ref{asmp-error} lead to consistent estimates for the common factors using the control units in the first stage of our estimation \citep{chen2021quantile}.

Assumption \ref{asmp-weakdep} posits the stationarity and weak dependence structure on the random shocks on the treated unit. Since $\epsilon_{1t}$ represents the modeling errors that are not captured by common factors, it is natural to assume that they follow this standard assumption about idiosyncratic shocks. 
Assumption \ref{asmp-weakdep} lends us the law of large numbers and the central limit theorem for quantile regression with time series data \citep{gregory2018smooth}, which is commonly used in the asymptotic analysis of synthetic control methods \citep{li2017estimation,cai2022estimating}. 
Assumption \ref{asmp-iden}(i) posits the strict exogeneity which is typical to identify quantile treatment effects \citep{firpo2007efficient}. It requires that there are nothing other than the common factors that jointly affect the outcome and treatment.
Assumption \ref{asmp-iden}(ii) establishes the uniqueness of the parameters $\beta_0=(\lambda_{01},\delta_0)$ in the quantile regression in Algorithm \ref{algo-QTE}. Denote $\Upsilon=\lim_{T\to\infty}T^{-1}\sum_{t=1}^T w_{0t}w_{0t}'$. The positive definiteness of $\Gamma$ holds if $\Upsilon$ is positive definite, $h_{1t}(0)$ is uniformly bounded away from zero and $h_{1t}(\epsilon)$ does not depend on $f_{0t}$ at $\epsilon=0$. 
Assumption \ref{asmp-iden}(iii) requires that we have enough observations in both pre-treatment and post-treatment periods to identify the quantile treatment effect.

Our first theoretical result is that the NQTT estimator $\widehat{\delta}$ based on Algorithm \ref{algo-IQR} (IQR) is consistent.

\begin{theorem}
\label{thm1}
    Under Assumptions \ref{asmp-factor}-\ref{asmp-iden}, we have $\widehat{\delta}\stackrel{p}{\to}\delta_0$.
\end{theorem}

As illustrated in the previous section, the estimated factor $\widehat{f}_t$ from Algorithm \ref{algo-IQR} may not achieve the desired convergence rate for $\widehat{\delta}$ to be asymptotically normal. However, $\widetilde{f}_t$ from Algorithm \ref{algo-ISQR} is able to attain such rate due to its smoothed objective function. To establish the asymptotic normality for $\widetilde{\delta}(\tau)$, we need some additional notations and assumptions.
Define $\psi_\tau(u)=\tau-\mathbf{1}(u<0)$, and
\begin{gather*}
    \Sigma= \lim_{T\to\infty}\sum_{s=-T}^{T}w_{0t}w_{0,t-s}'\mathbb{E}[\psi_\tau(\epsilon_{1t})\psi_\tau(\epsilon_{1,t-s})|f_{0t},f_{0,t-s}], \\
    \Phi_i=\lim_{T\to\infty}\frac{1}{T}\sum_{t=1}^Th_{it}(0)f_{0t}f_{0t}',\,\, \Psi_t=\lim_{N\to\infty}\frac{1}{N}\sum_{i=2}^{N+1} h_{it}(0)\lambda_{0i}\lambda_{0i}'
\end{gather*}
\begin{assumption}[Asymptotic Variance]
\label{asmp-var}
$\Sigma$ is positive definite.
\end{assumption}

\begin{assumption}[Smoothed Quantile Regression]
\label{asmp-smooth}
Let $m\geq 8$ be a positive integer.
\begin{enumerate}[(i)]
    \item $\Phi_i$ and $\Psi_t$ are positive definite for all $i\geq  2,t\geq 1$.
    \item $\lambda_{0i}$ is an interior point of $\mathcal{A}$ and $f_{0t}$ is an interior point of $\mathcal{F}$ for all $i\geq  2,t\geq 1$.
    \item $k(z)$ is symmetric around $0$ and twice continuously differentiable. $\int_{-1}^1 k(z)dz=1$, $\int_{-1}^1 z^jk(z)dz=0$ for $j=1,...,m-1$ and $\int_{-1}^1 z^mk(z)dz\neq 0$.
    \item $h_{it}(\epsilon)$ is $m+2$ times continuously differentiable. Let $h_{it}^{(j)}(\epsilon)=(\partial/\partial\epsilon)^j h_{it}(\epsilon)$ for $j=1,...,m+2$. For any compact set $C\subset\mathbb{R}$ and any $\epsilon\in C$, there exist $-\infty<\underline{l}<\overline{l}<\infty$ (depending on $C$) such that $\underline{l}\leq h_{it}^{(j)}(\epsilon)\leq \overline{l}$ and $\underline{h}\leq h_{it}(\epsilon)\leq \overline{l}$ for $j=1,...,m+2$ and for all $i\geq 2,t\geq 1$.
    \item As $N,T\to\infty$, $N\propto T$, $h\propto T^{-c}$ and $m^{-1}<c<1/6$.
\end{enumerate}
\end{assumption}

Assumption \ref{asmp-iden} requires $\Sigma$, a component of the asymptotic variance of $\widetilde{\delta}(\tau)$, to be positive definite. $\Sigma$ is the long-run variance-covariance structure of $\epsilon_{1t}$ that contains the temporal dependence in the data \citep{davidson1994stochastic}. If $\epsilon_{1t}$ is uncorrelated across $t$, then $\Sigma$ equals $\tau(1-\tau)\Upsilon$ as the quantile regression for i.i.d. data.
Assumption \ref{asmp-smooth} resembles Assumption 2 in \cite{chen2021quantile}, which is standard in smoothed quantile regressions \citep{galvao2016smoothed}. Assumption \ref{asmp-smooth}(i) is a similar version of Assumption \ref{asmp-iden}(ii) for the control units. Assumption \ref{asmp-smooth}(ii) is mild once we properly define the parameter space $\mathcal{A}$ and $\mathcal{F}$. Assumption \ref{asmp-smooth}(iii) is a technical requirement for high-order kernels used in the smoothed objective function and many candidates are available in the literature \citep{li2007nonparametric}. Assumption \ref{asmp-smooth}(iv) demands more smoothness for conditional density functions. Assumption \ref{asmp-smooth}(v) states that $N$ and $T$ should grow at the same rate, rendering a balanced panel, and regulates the rate of bandwidth $h$ of the kernel.

\begin{theorem}
\label{thm2}
    Under Assumptions \ref{asmp-factor}-\ref{asmp-smooth}, we have
    \begin{align} \label{dist_thm2}
        \sqrt{T}(\widetilde{\delta}-\delta_0)\stackrel{d}{\to} \mathcal{N}(0,\Omega),
    \end{align}
    as $N$ and $T$ $\rightarrow \infty$, where $\Omega$ is the $(r+1)(r+1)$-th element of $\Gamma^{-1}\Sigma\Gamma^{-1}$.
\end{theorem}
Theorem \ref{thm2} enables pointwise inference for the SQTT estimator $\widetilde{\delta}$, e.g., the confidence interval for each $\tau$-th quantile as shown in Figure \ref{fig1} and \ref{fig2} in Section \ref{sec:empirical} for some given $\tau\in(0,1)$. It would be interesting to examine the uniform inference for an interval of quantiles $\mathcal{T}=[\tau_1,\tau_2]\subset (0,1)$, which allows researchers to test global hypotheses about the entire distribution of potential outcomes \emph{en bloc}, and facilitates comparisons between different quantiles. Establishing uniform inference for $\widetilde{\delta}$ might require the use of empirical process theory \citep{vaart2023empirical} to deal with the estimated regressors $\widetilde{f}_t$ in the quantile regression process $\{\sqrt{T}(\widetilde{\delta}(\tau)-\delta_0(\tau)),\tau\in\mathcal{T}\}$, which is essentially different from proving the pointwise asymptotic distribution \citep{angrist2006quantile}. We leave this for the future research avenue.

Remarkably, the asymptotic properties of our QTT estimators are not affected when the number of factors is consistently estimated as in Algorithm \ref{algo-MSRM}. Specifically, \cite{chen2021quantile} show that $\lim_{N,T\rightarrow\infty} \mathbb{P}[\widehat{r}_{RM}=r_0]=1$. Following the argument in \citet[p.143]{bai2003inferential}, for $\widecheck{f}_t$ being $\widehat{f}_t$ or $\widetilde{f}_t$, we have $\mathbb{P}[\widecheck{f}_t\leq z]=\mathbb{P}[\widecheck{f}_t\leq z,\hat{r}_{RM}=r_0]+\mathbb{P}[\widecheck{f}_t\leq z,\hat{r}_{RM}\neq r_0]$. Observe that $\mathbb{P}[\widecheck{f}_t\leq z,\widehat{r}_{RM}\neq r_0]\leq \mathbb{P}[\widehat{r}_{RM}\neq r]=o(1)$. Thus, 
\begin{align}
    \nonumber \mathbb{P}[\widecheck{f}_t\leq z] &=\mathbb{P}[\widecheck{f}_t\leq z,\widehat{r}_{RM}=r_0]+o(1) \\
    \nonumber &=\mathbb{P}[\widecheck{f}_t\leq z|\widehat{r}_{RM}=r_0]\mathbb{P}[\widehat{r}_{RM}=r_0] + o(1) \\
    &=\mathbb{P}[\widecheck{f}_t\leq z|\widehat{r}_{RM}=r_0] + o(1),
\end{align}
since $\mathbb{P}[\widehat{r}_{RM}=r_0]\to 1$. Therefore, asymptotic properties of the estimated factors $\widecheck{f}_t$ using $\widehat{r}_{RM}$ are the same as using $r_0$, and so are our QTT estimators.

The last part of this section is to establish the bootstrap validity for Algorithm \ref{algo-blockboot}. We need an additional assumption for the bootstrap block size.

\begin{assumption}
\label{asmp-boot}
    The boostrap block size $\mathscr{B}_d$ satisfies $\mathscr{B}_d^{-1}+\mathscr{B}_d/T_d\to 0$ as $T\to \infty$ for $d=0,1$.
\end{assumption}
Assumption \ref{asmp-boot} simply requires that the block size increases with $T$ at a slow enough rate, which is a mild restriction. The next theorem shows that the block boostrap procedure in Algorithm \ref{algo-blockboot} is consistent for the asymptotic distribution in Theorem \ref{thm2} \citep{horowitz2019bootstrap}, so that the bootstrap confidence interval is valid.

\begin{theorem}
\label{thm3}
    Let $\widetilde{\delta}^\ast$ be the bootstrap version of $\widetilde{\delta}$ from Algorithm \ref{algo-blockboot}. Under Assumptions \ref{asmp-factor}-\ref{asmp-boot}, we have as $N,T\to\infty$,
    \begin{align}        \sup_{z\in\mathbb{R}^{r+1}}\bigg|\mathbb{P}^\ast\big[\sqrt{T}(\widetilde{\delta}^\ast-\widetilde{\delta})\leq z\big] - \mathbb{P}\big[\sqrt{T}(\widetilde{\delta}-\delta_0)\leq z\big]\bigg|\stackrel{p}{\to} 0,
    \end{align}
    where $\mathbb{P}^\ast$ denotes bootstrap probability.
\end{theorem}

\section{Simulation}
\label{sec:simulation}

In this section, we report the finite sample performance of our method using Monte-Carlo experiments. The sample sizes are set to be $N+1 \in \{51,101,201\}$ and $T \in \{100,200,400\}$. A treatment starts to affect the outcome of interest $y$ at period $T_0 = T/2+1$. We consider a DGP for the potential outcome in the absence of treatment following \cite{chen2021quantile}:
\begin{align} \label{simeq}
    y^0_{it} = \lambda_{1i}f_{1t} + \lambda_{2i}f_{2t}  + \lambda_{3i}f_{3t} u_{it}, 
\end{align}
where $f_{1t}=0.8f_{1,t-1}+\nu_{1t}$, $f_{2t}=0.5f_{2,t-1}+\nu_{2t}$, $f_{3t}=|g_{t}|$, $\lambda_{1i}, \lambda_{2i}, \nu_{1t}, \nu_{2t}, g_{t}, u_{it}\stackrel{\text{i.i.d}}{\sim}\mathcal{N}(0,1)$, and $\lambda_{3i}\stackrel{\text{i.i.d}}{\sim}\mathcal{U}(1,2)$. The observed outcome is the following:

\begin{equation}
y_{it} =\left\{
\begin{aligned}
&y_{it}^0 + \alpha_{t}, \hspace{0.8cm} t > T_0, i=1; \\
&y_{it}^0, \hspace{1.7cm} \text{otherwise}.
\end{aligned}
\right.
\end{equation}
We set $\alpha_{t} = u_{1t} + 0.5$ so that the true quantile treatment effect is $\delta_{0}(\tau) = 0.5 + \Phi^{-1}(\tau)$, where $\Phi$ is the CDF function of the standard normal distribution. One may fit \eqref{simeq} into QFM by writing $\lambda_i(\tau)=(\lambda_{1i},\lambda_{2i})'$, $f_t(\tau) = (f_{1t},f_{2t})'$ if $\tau= 0.5$, and $\lambda_i(\tau)=(\lambda_{1i},\lambda_{2i},\lambda_{3i} \Phi^{-1}(\tau))'$, $f_t(\tau) = (f_{1t},f_{2t},f_{3t})'$ if $\tau\neq0.5$.

We compare the proposed estimators with two other estimation results, where \emph{NQTT} (Non-smoothed QTT) and \emph{SQTT} (Smoothed QTT) denote the estimators $\widehat{\delta}$ and $\widetilde{\delta}$ following Algorithm \ref{algo-IQR} and \ref{algo-ISQR}, respectively. In \emph{SQTT}, we use the kernel function suggested by \cite{chen2021quantile}\footnote{$k(z)=\mathbf{1}(|z|<1)\frac{3465}{8192} \left(7 - 105 z^{2} + 462 z^{4} - 858 z^{6} + 715 z^{8} - 221 z^{10} \right)$}, and set $h=0.5$ (since $h$ would not change much with $T$ given the rate in Assumption \ref{asmp-smooth}(v)). \emph{Oracle} denotes the direct quantile regression had we observed the complete set of factors $f_t$, and \emph{GSCM} denotes the general synthetic control method of \cite{xu_2017}. While \emph{Oracle} simplifies to a standard quantile regression estimation, \emph{GSCM} first estimates the factor number $\widehat{r}_{IC}$ based on the information criteria considered in \cite{bai2002determining}\footnote{$\widehat{r}_{IC} = \argmin_{r \in \{2,\dots,5\}}  \text{IC}(r) = \argmin_{r \in \{2,\dots,5\}} \ln \left( \frac{1}{NT} \sum_{i=2}^{N+1}\sum_{t=1}^T (y_{it}-\widehat{\lambda}_i'\widehat{f}^r_{t,PCA})^2 \right)+ r\frac{N+T}{NT} \ln ( \frac{NT}{N+T} ).$}, and then uses principle component analysis (PCA) to estimate the factor $\widehat{f}^r_{t, PCA}$, given $r=\widehat{r}_{IC}$. Finally, for each quantile level $\tau$, \emph{GSCM} runs the standard quantile regression with $\widehat{f}^r_{t, PCA}$ and an indicator. The coefficient of the indicator is considered as the estimated treatment effect. Note that $\widehat{f}^r_{t, PCA}$ remains the same across $\tau$, which is different from our two QTT estimators.\footnote{We note that \emph{GSCM} is simplified to PCA in our simulations, since we do not consider the existence of covariates. When covariates are included, \emph{GSCM} relies on an iterative procedure to estimate the covariate coefficients, factors, and loadings.}

First, we assess the performance of the estimators by computing the bias and rooted mean square error (RMSE) at $\tau\in\{0.1,0.25,0.5,0.75,0.9\}$ across $R=1000$ simulations:
\begin{align}
    & \operatorname{Bias}_{m}(\tau) = \frac{1}{R} \sum_{r=1}^R (\delta(\tau)_{r,m}-\delta_{0}(\tau)),\,\,
    \operatorname{RMSE}_{m}(\tau) = \bigg[\frac{1}{R} \sum_{r=1}^R(\delta(\tau)_{r,m}-\delta_{0}(\tau))^2\bigg]^{1/2}, \notag
\end{align}
where $m \in \{NQTT, SQTT, Oracle, GSCM\}$.
Table \ref{tab-1} presents the result. We see that \emph{NQTT} and \emph{SQTT} both produce consistent estimates across all quantiles compared to \emph{Oracle}. In contrast, \emph{GSCM} exhibits severe biases at $\tau\neq0.5$, as \emph{GSCM} could not capture $f_{3t}$ and thus fails to recover the true quantile structure in this case.

\begin{table}[!ht]
\centering
\caption{Bias and RMSE}
\label{tab-1}
\begin{tabular}{cccccccccc}
  \toprule
   \multicolumn{3}{c}{$\tau$} & 0.1 & 0.25 & 0.5 & 0.75 & 0.9  \\ 
   
  \midrule

\multirow{8}{*}{\makecell{N=50 \\ T=100}} 
& \multirow{2}{*}{NQTT} & Bias &0.1357 & -0.1671 & -0.0128 & 0.1802 & -0.0760   \\
& & RMSE &0.4964 & 0.4593 & 0.3069 & 0.4711 & 0.5110   \\ \cmidrule{2-8}

& \multirow{2}{*}{SQTT} & Bias & 0.0769 & -0.2318 & 0.0009 & 0.2459 & -0.0753      \\
& & RMSE & 0.4959 & 0.5018 & 0.3116 & 0.5137 & 0.5058    \\ 
\cmidrule{2-8}

& \multirow{2}{*}{Oracle} & Bias & 0.0866 & 0.0189 & -0.0092 & -0.0160 & -0.0337     \\
& & RMSE &0.4304 & 0.3591 & 0.3311 & 0.3591 & 0.4352    \\ 
\cmidrule{2-8}

& \multirow{2}{*}{GSCM} & Bias &-1.4205 & -0.6215 & -0.0097 & 0.6058 & 1.4279   \\
& & RMSE &1.5593 & 0.7450 & 0.3398 & 0.7371 & 1.5697  \\
\midrule

\multirow{8}{*}{\makecell{N=100 \\ T=200}} 
& \multirow{2}{*}{NQTT} & Bias &0.0646 & -0.0461 & -0.0010 & 0.0707 & -0.0402    \\
& & RMSE &0.4192 & 0.3569 & 0.2605 & 0.3506 & 0.3976   \\ \cmidrule{2-8}

& \multirow{2}{*}{SQTT} & Bias &  0.0456 & -0.0689 & -0.0037 & 0.0558 & -0.0561 \\  
& & RMSE & 0.3956 & 0.3627 & 0.2562 & 0.3528 & 0.4117   \\ 
\cmidrule{2-8}

& \multirow{2}{*}{Oracle} & Bias &0.0396 & 0.0195 & 0.0138 & -0.0011 & -0.0077     \\
& & RMSE &0.3797 & 0.2954 & 0.2714 & 0.2959 & 0.3686   \\ \cmidrule{2-8}

& \multirow{2}{*}{GSCM} & Bias &-1.8524 & -0.7761 & 0.0006 & 0.7799 & 1.8244  \\
& & RMSE &1.9392 & 0.8401 & 0.2643 & 0.8442 & 1.8998    \\
\midrule

\multirow{8}{*}{\makecell{N=200 \\ T=400}} 

& \multirow{2}{*}{NQTT} & Bias &0.0255 & -0.0094 & -0.0008 & 0.0062 & -0.0189   \\
& & RMSE &0.2283 & 0.1802 & 0.1633 & 0.1903 & 0.2295  \\ \cmidrule{2-8}

& \multirow{2}{*}{SQTT} & Bias & 0.0246 & -0.0265 & -0.0008 & 0.0144 & -0.0335  \\
& & RMSE &0.2731 & 0.2231 & 0.1815 & 0.2201 & 0.2719  \\ \cmidrule{2-8}

& \multirow{2}{*}{Oracle}  & Bias   &0.0189 & 0.0034 & -0.0031 & -0.0053 & -0.0121    \\
& & RMSE  &0.2196 & 0.1808 & 0.1741 & 0.1908 & 0.2271   \\
\cmidrule{2-8}

& \multirow{2}{*}{GSCM} & Bias &-1.3006 & -0.5776 & -0.0007 & 0.5749 & 1.3084   \\
& & RMSE &1.3309 & 0.6052 & 0.1640 & 0.6062 & 1.3429    \\
\bottomrule

\end{tabular}
\end{table}

Second, we validate our inference theory by reporting the standard deviation and coverage probability for the 95\% confidence intervals using the blockwise bootstrap procedure. The bootstrap sample size is $B=1000$. The standard deviation and coverage probabilities are calculated as follows:
\begin{align}
    & \operatorname{SD}(\tau) = \frac{1}{R} \sum_{r=1}^R \widecheck{\sigma}_r^\ast(\tau), \,\,
    \operatorname{Coverage}(\tau) =  \frac{1}{R} \sum_{r=1}^R \mathbf{1}\{\widecheck{\delta}_r(\tau) \in (\widecheck{\delta}_r^\ast(\tau)-1.96\widecheck{\sigma}_r^\ast,\widecheck{\delta}_r^\ast(\tau)+1.96\widecheck{\sigma}_r^\ast)\}, \notag
\end{align}
where the estimator $\widecheck{\delta}$ is either NQTT $\widehat{\delta}(\tau)$ or SQTT $\widetilde{\delta}(\tau)$. Note that although we only provide the asymptotic normality and bootstrap validity theory for SQTT (Theorem \ref{thm2} and \ref{thm3}), we also conduct the same bootstrap inference for NQTT in this simulation to examine its finite sample performance.
Table \ref{tab-2} shows that for both NQTT and SQTT, the SD declines as the sample size increases and the accuracy of the coverage probability is satisfactory. This demonstrates the effectiveness of our asymptotic theory for SQTT, and suggests that valid inference might also hold for NQTT. 

\begin{table}[!t] 
\centering
\caption{Standard Deviation and 95\% Coverage for Bootstrap}
\label{tab-2}
\begin{tabular}{ccccccccc}
  \toprule
 &  \multicolumn{2}{c}{$\tau$} & 0.1 & 0.25 & 0.5 & 0.75 & 0.9  \\ 
   
  \midrule

\multirow{4}{*}{\makecell{N=50 \\ T=100}}  

& \multirow{2}{*}{NQTT} & SD   &0.5448 & 0.3873 & 0.3182 & 0.3934 & 0.5474  \\   
& & Coverage &0.9520 & 0.8800 & 0.9540 & 0.9090 & 0.9440  \\ \cmidrule{2-8}

& \multirow{2}{*}{SQTT} & SD   &0.5523 & 0.3915 & 0.3208 & 0.3890 & 0.5433  \\   
& & Coverage &0.9470 & 0.8810 & 0.9430 & 0.8710 & 0.9450  \\ 

\midrule

\multirow{4}{*}{\makecell{N=100 \\ T=200}}  

& \multirow{2}{*}{NQTT} & SD   & 0.4222 & 0.3240 & 0.2636 & 0.3265 & 0.4183 \\    
& & Coverage &0.9500 & 0.9310 & 0.9440 & 0.9520 & 0.9660 \\  \cmidrule{2-8}

& \multirow{2}{*}{SQTT} & SD   &0.4169 & 0.3219 & 0.2604 & 0.3217 & 0.4201  \\   
& & Coverage &0.9570 & 0.9290 & 0.9470 & 0.9310 & 0.9500  \\ \midrule

\multirow{4}{*}{\makecell{N=200 \\ T=400}}  

& \multirow{2}{*}{NQTT} & SD   &0.2396 & 0.1848 & 0.1611 & 0.1885 & 0.2425   \\
& & Coverage &0.9610 & 0.9540 & 0.9470 & 0.9450 & 0.9540  \\ \cmidrule{2-8}

& \multirow{2}{*}{SQTT} & SD   &0.2872 & 0.2242 & 0.1887 & 0.2252 & 0.2902   \\   
& & Coverage &0.9510 & 0.9480 & 0.9530 & 0.9520 & 0.9640  \\ 
\bottomrule

\end{tabular}
\end{table}

In the Appendix, we conduct further simulations to safeguard our estimators under different DGP setups, including heavy-tailed errors (allowed by our theory), serially and cross-sectionally correlated errors (not allowed by our theory), and more complicated factor structures (allowed by our theory). We arrive at similar results as the above baseline simulation in which our two proposed estimators consistently recover the true quantile treatment effects and permit valid bootstrap inference.

\section{Empirical Application}
\label{sec:empirical}

In this section, we apply our proposed method to evaluate the effects of the 2008 Chinese Economic Stimulus Program.\footnote{We revisit California’s tobacco control program \citep{abadie2010synthetic} in Section \ref{sec:cali} in the Appendix.} At the end of 2008, the Chinese government launched an economic stimulus package amounting to four trillion RMB (equivalent to 586 billion USD) as a response to the global financial crisis, with the aim of counteracting its adverse impacts on the domestic economy. 
Assessing the effectiveness of this fiscal intervention will deepen our understanding on large-scale economic stimulus programs.

The economic outcomes following the policy’s implementation are influenced by various latent factors, such as global financial crises and trade dynamics, alongside the stimulus itself. Recognizing these complexities, \cite{ouyang2015treatment} improve the SCM approach originally proposed by \cite{hsiao2012panel}, which is grounded in a factor model structure, to estimate the ATT of the stimulus. 
Their findings revealed that the fiscal stimulus program boosted China’s annual real GDP growth by approximately 3.2\% on average.

Although the ATT provides valuable insights, policymakers might also be interested in how the stimulus program influenced the distribution of macroeconomic outcomes. Specifically, examining the QTT offers a deeper understanding of the policy impact under varying economic conditions. For instance, QTT at the 90\% quantile sheds light on the policy’s effectiveness in optimistic scenarios, while QTT at lower quantiles highlights its role during economic downturns. Next, we apply our QTT estimators to explore the impact of the stimulus program at different quantiles of GDP growth and investment.

\subsection{Data Description}\label{sec:data-pretrend}

To evaluate QTT, we collect quarterly real GDP growth data from the OECD statistical databases, following the same data source as \cite{ouyang2015treatment}, but with an expanded dataset. The control group comprises 40 countries, and the GDP data spans from the first quarter of 1999 to the fourth quarter of 2015, giving a pre-treatment period of $T_0 = 40$ quarters and a post-treatment period of $T_1 = 28$ quarters. The real GDP growth is measured as the annual growth rate—the difference between the GDP level of the current quarter and the same quarter of the previous year—to ensure stationarity and eliminate seasonality.

\begin{figure}[!t]
\centering
\caption{Pre-trend of GDP Growth Rate}
\includegraphics[scale=0.45]{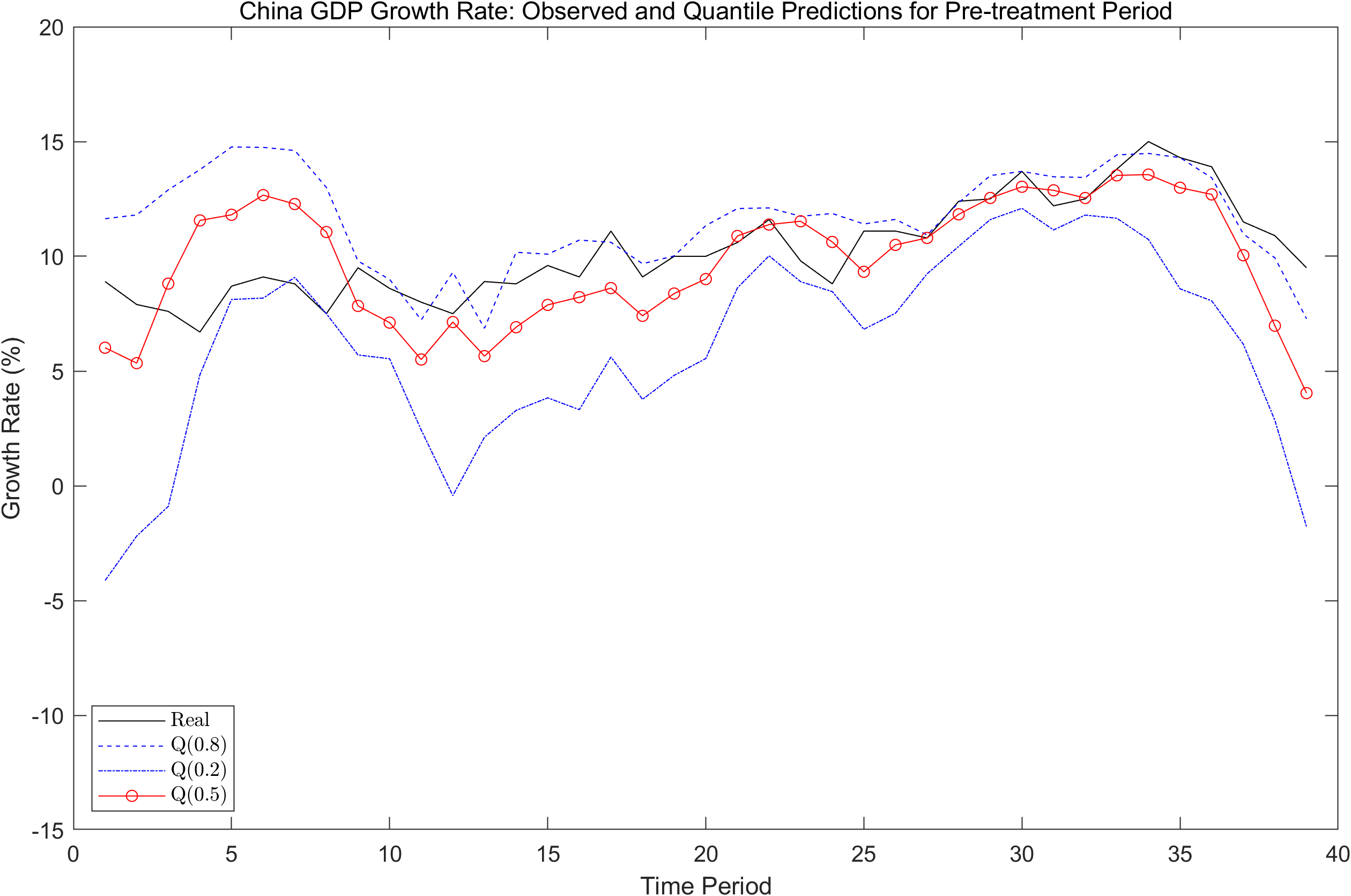}
\caption*{Notes: The black solid line represents the observed China GDP growth rate, while the upper and lower blue dashed lines represent the predicted $80\%$ and $20\%$ quantiles at each time period, i.e., $\widehat{Q}_{0.8}(y_{t}|\widehat{f}_t)$ and $\widehat{Q}_{0.2}(y_{t}|\widehat{f}_t)$, respectively. The red dotted line is the predicted median $\widehat{Q}_{0.5}(y_{t}|\widehat{f}_t)$. The time line is the whole pre-treatment periods, $t=1$ to $40$.}
\label{2008gdp_pretrend}
\end{figure}

The pre-treatment trend of GDP growth rate is described in Figure \ref{2008gdp_pretrend}. We plot the actual China GDP growth rate together with the quantile predictions $\widehat{Q}_\tau(y_{1t}|\widehat{f}_t)$, where the estimated factors $\widehat{f}_t$ is from QFM using the control group. We see that the median prediction closely traces the observed outcome, suggesting that the estimated factors from the control group explain the treated unit fairly well. This aligns with the convex hull condition \citep{abadie2021using} in SCM that a combination of unaffected units can approximate
the pre-treated outcome of the affected unit. Our QTT, for example, at $80\%$ quantile, is to uncover how $\widehat{Q}_{0.8}(y_{1t}|\widehat{f}_t)$ would change after the policy. The difference between observed outcome and the quantile predictions, i.e., $y_{1t}-\widehat{Q}_\tau(y_{1t}|\widehat{f}_t)$, is just the (estimated) idiosyncratic error $\widehat{\epsilon}_{it}(\tau)$, which may reflect the random shock unexplained by latent factors.

\subsection{Policy Impact on Real GDP Growth}

\begin{figure}[!t]
  \centering
  \caption{2008 China Stimulus Package on GDP Growth Rate}
  \begin{subfigure}[b]{0.45\textwidth-\columnsep}
    \includegraphics[width=\textwidth, height=0.3\textheight, keepaspectratio]{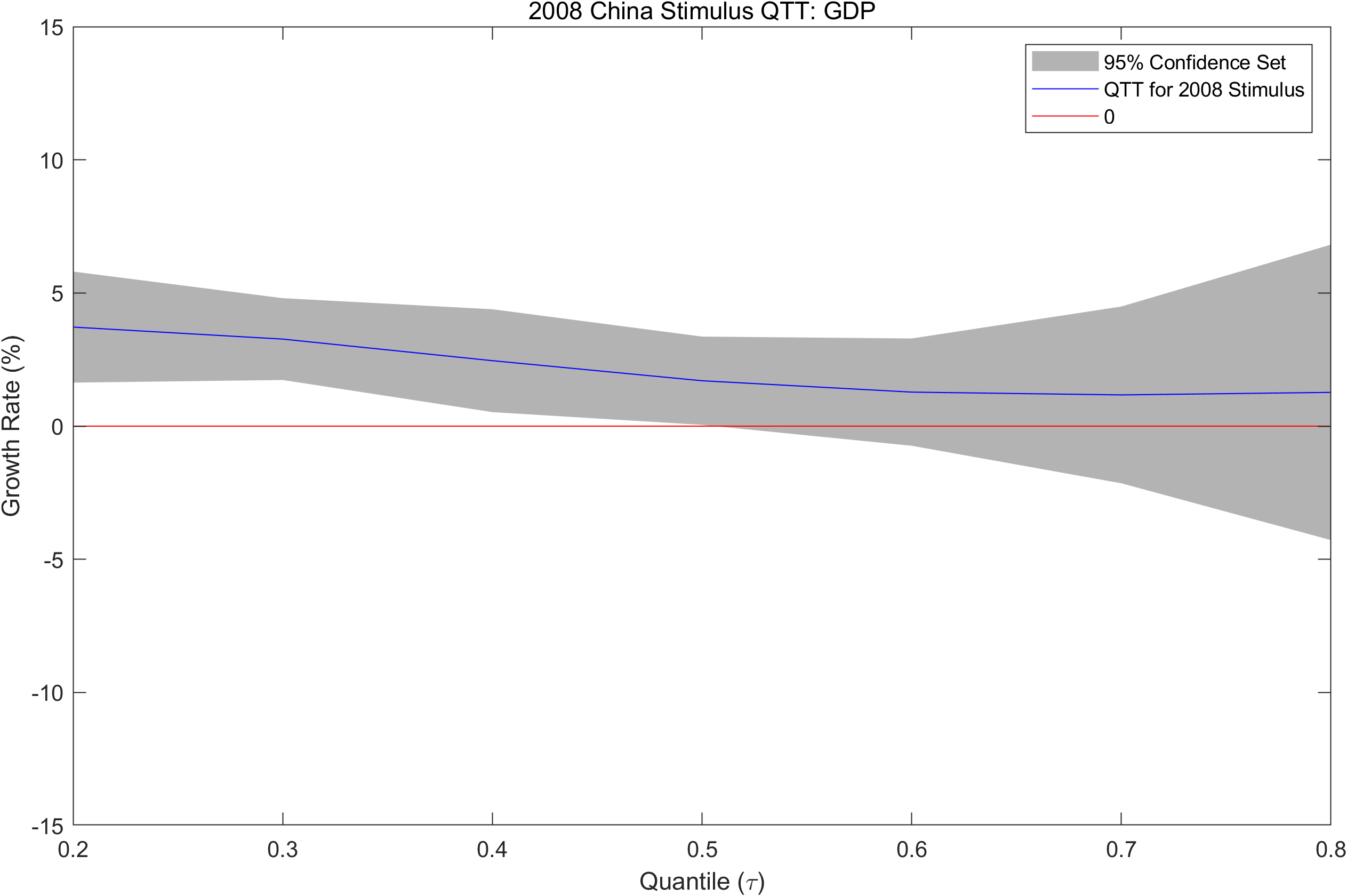}
    \caption{NQTT}
    \label{2008sti_gdp}
  \end{subfigure}
  \hfill 
  \begin{subfigure}[b]{0.45\textwidth-\columnsep}
    \includegraphics[width=\textwidth, height=0.3\textheight, keepaspectratio]{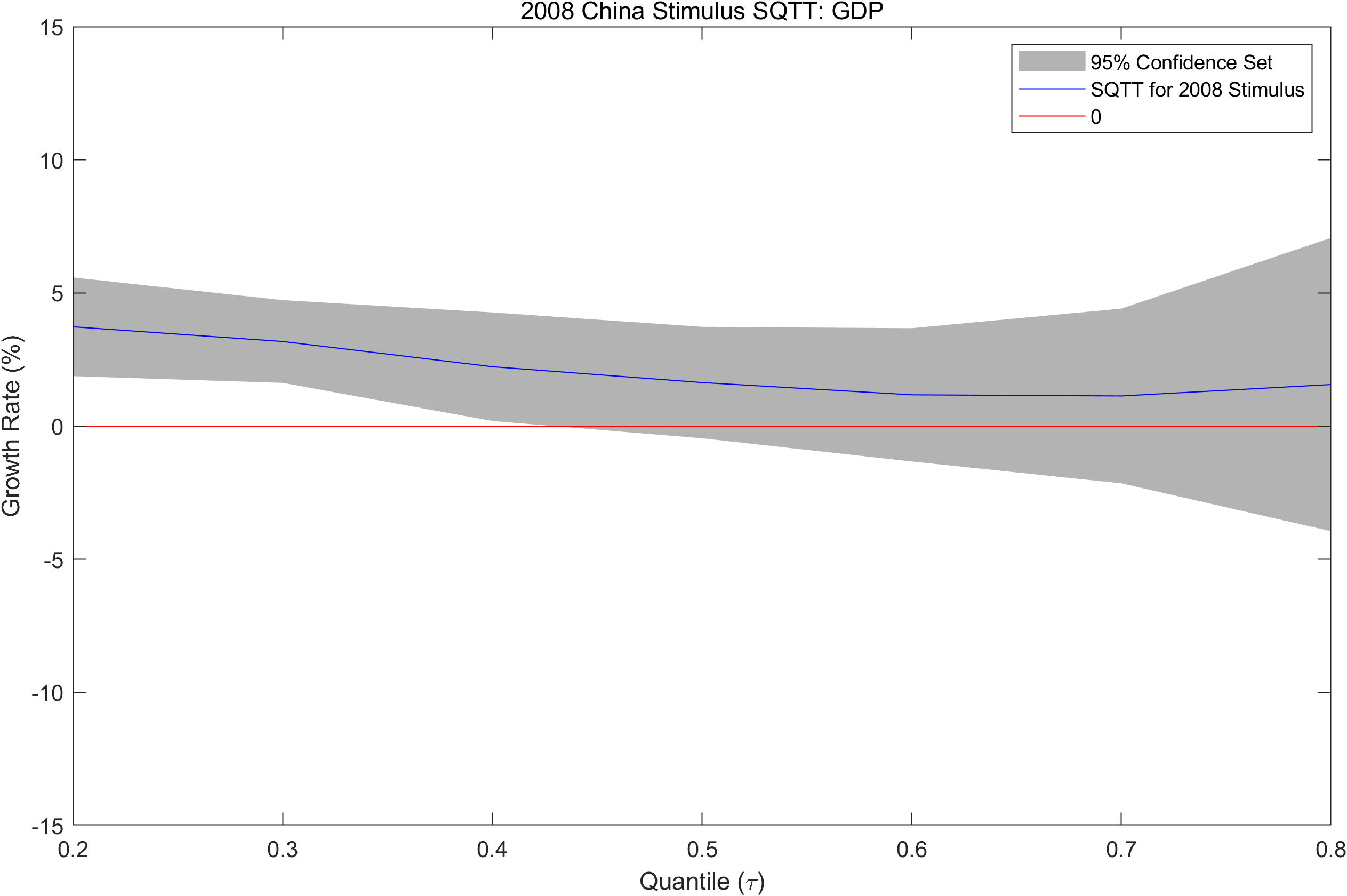}
    \caption{SQTT}
    \label{2008sti_gdp_SQTT}
  \end{subfigure}
  \caption*{Note: The blue line is the estimated QTT, and the shaded area represents the $95\%$ confidence interval.}
  \label{fig1}
\end{figure}

We apply our QTT estimators to study the impact of the 2008 economic stimulus program on China's real GDP growth. Figure \ref{fig1} presents the estimated QTTs across different quantiles, along with 95\% confidence intervals based on blockwise bootstrap with $1000$ replications.

The results are consistent between \emph{NQTT} and \emph{SQTT}, indicating that the stimulus program had significantly positive effects on GDP growth at quantiles below 50\%, suggesting that the policy served as a safety net to sustain economic growth during adverse global economic conditions. However, for the quantiles above 50\%, while the estimated effects remain positive, they are not statistically significant. This highlights the program’s primary function as a stabilizer rather than as a mechanism to maximize economic growth in favorable scenarios. The findings underscore that fiscal stimulus policies are more effective in mitigating economic downturns than boosting growth ceilings.

\subsection{Policy Impact on Gross Fixed Capital Formation}

We further examine the influence of the stimulus program on one of the key components of real GDP: Gross Fixed Capital Formation (investment). \cite{ouyang2015treatment} report that the stimulus program increased China’s real investment growth by 22.15\% and note that investment responded with the fastest and largest magnitude among the economic components.

\begin{figure}[!t]
  \centering
  \caption{2008 China Stimulus Package on Gross Fixed Capital Formation}
  \begin{subfigure}[b]{0.45\textwidth-\columnsep}
    \includegraphics[width=\textwidth, height=0.3\textheight, keepaspectratio]{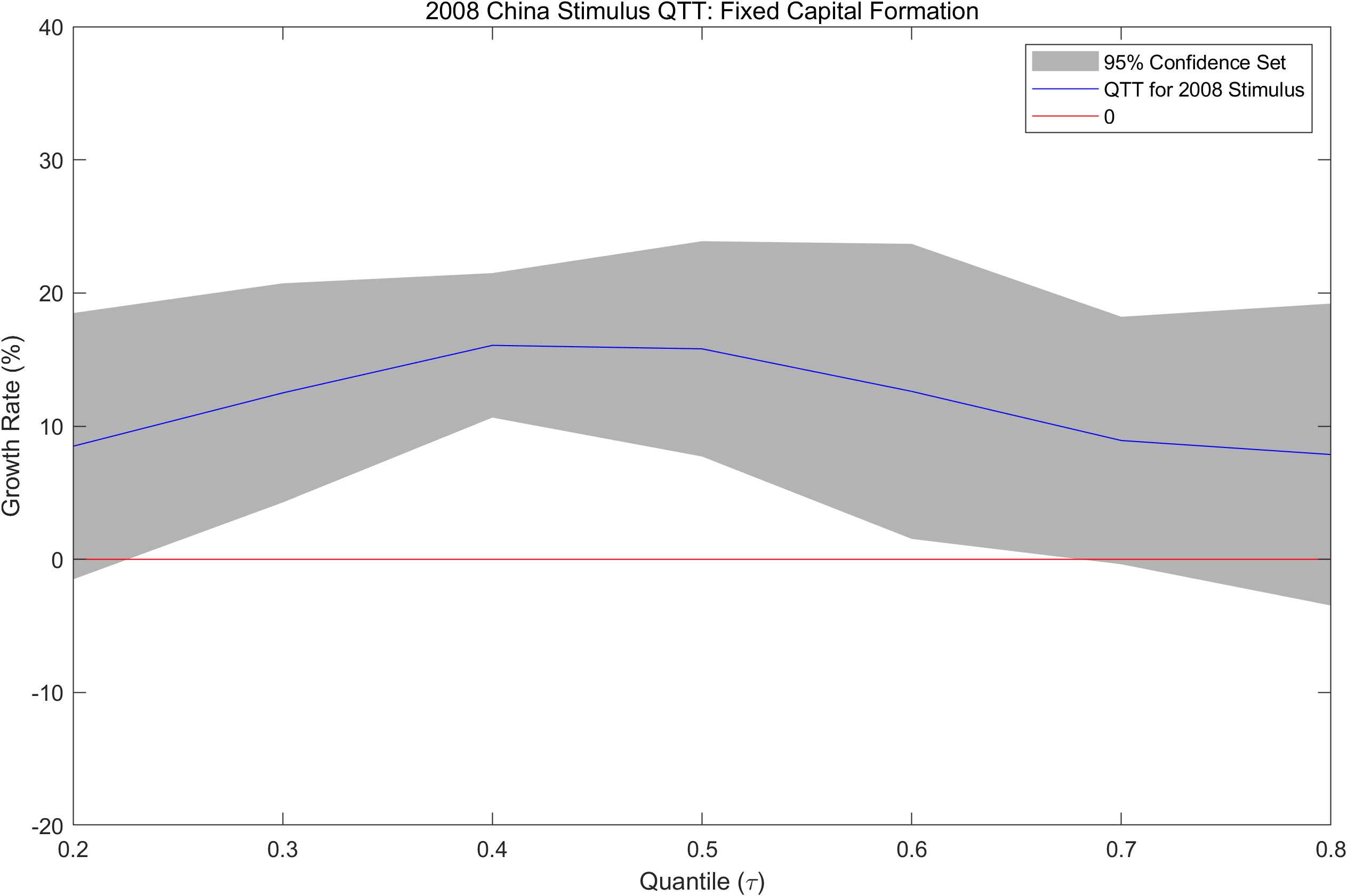}
    \caption{NQTT}
    \label{2008sti_invest}
  \end{subfigure}
  \hfill 
  \begin{subfigure}[b]{0.45\textwidth-\columnsep}
    \includegraphics[width=\textwidth, height=0.3\textheight, keepaspectratio]{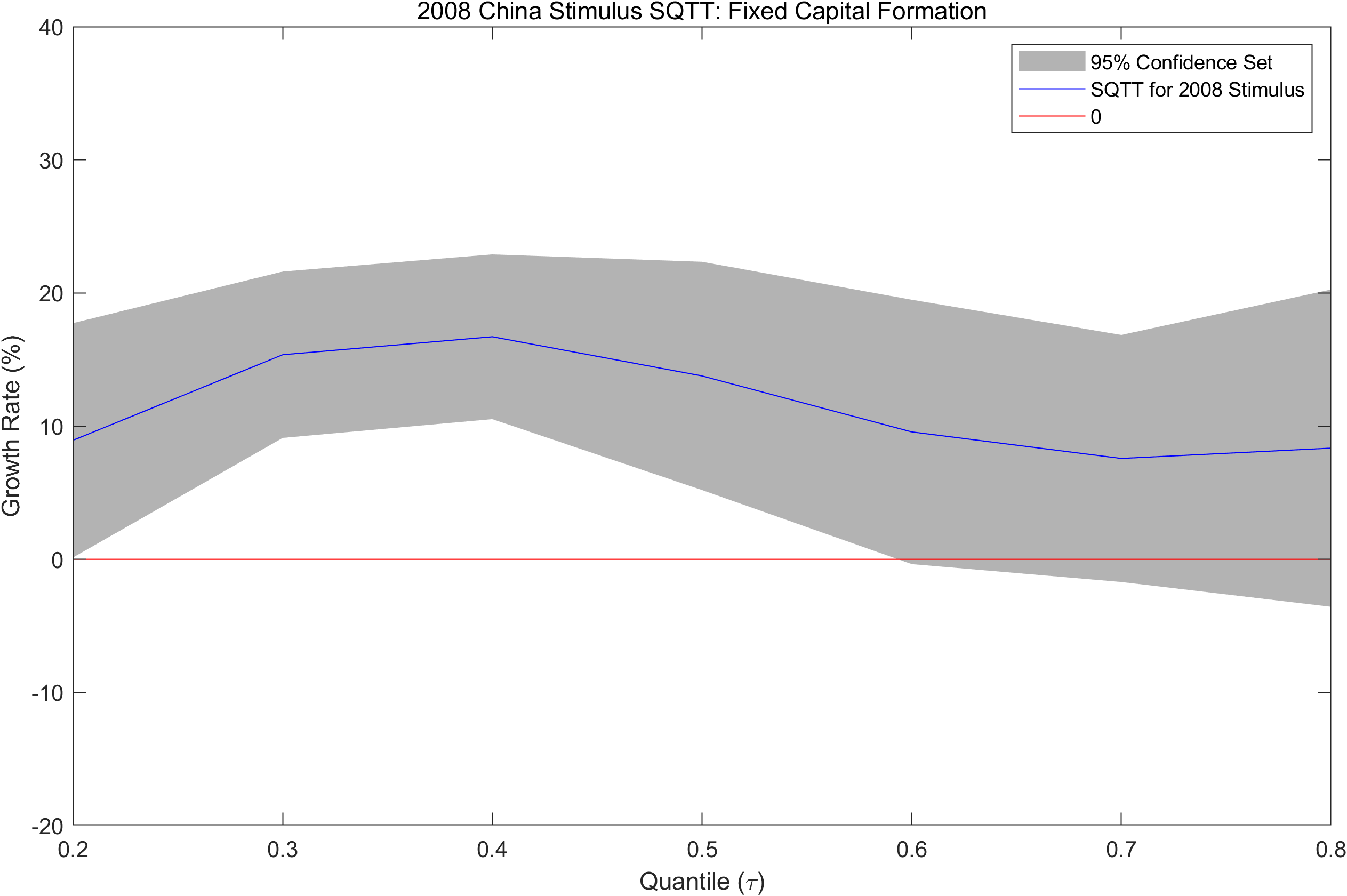}
    \caption{SQTT}
    \label{2008sti_invest_SQTT}
  \end{subfigure}
  \caption*{Note: The blue line is the estimated QTT, and the shaded area represents the $95\%$ confidence interval.}
  \label{fig2}
\end{figure}

Our analysis in Figure \ref{fig2} uncovers an intriguing pattern: the QTTs are significant for quantiles between 30\% and 50\%, where investment growth significantly outpaces GDP growth. This suggests that the influence of the stimulus program on investment is most pronounced under relatively bad economic conditions, while its effects diminish in other scenarios: either favorable or highly unfavorable. Moreover, the finding also indicates that the fiscal stimulus may skew resource allocation excessively toward investment at the expense of other GDP components. This imbalance highlights the potential unintended consequences of stimulus measures and underscores the importance of carefully considering their broader economic implications in future policy designs.

As a conclusion, the empirical findings provide a comprehensive view of the 2008 Chinese Economic Stimulus Program’s impacts. The analysis of QTTs reveals that the stimulus effectively stabilized economic growth and bolstered investment under economic downturns, potentially at the cost of directing resources toward investment and suppressing the growth of other GDP components. Such findings reinforce the view that fiscal stimulus should be considered as a temporary measure, and policymakers should carefully weigh its unintended consequences.

\section{Extensions}
\label{sec:extension}
In this section, we discuss some possible extensions of our method. As before, we omit the subscript $\tau$ for $\delta$ and $f_t$ whenever it does not cause confusion.

\subsection{Time Varying QTT}\label{sec:relax-linear}

Our model \eqref{qfs_linear} implies a constant QTT
$\delta = Q_{\tau}(y^1_{it} | f_t) - Q_{\tau}(y^0_{it} | f_t)$ for a given $\tau$. In many application, the policy might only exert transient impact so that it is more reasonable to consider a treatment effect that gradually fades out over time. Therefore, we discuss two extensions of our model in which the QTT is allowed to vary over time.

\subsubsection{QTT as A Function of Time}\label{sec:time-variant}

We could consider QTT as a function of time, i.e., $\delta_t = g(t)$, where $g(\cdot)$ is a deterministic function. For instance, one may let $g(t)=\zeta_0 + \zeta_1 / (t - T_0)$ for all $t \geq T_0 + 1$, which describes the idea that QTT gradually decays as time increases. Alternatively, if the researcher believes the intervention gradually builds up to a peak within a certain period after $T_0$ and subsequently diminishes, $g(t)$ can be modeled as a quadratic function of $t$.

We could directly extend the estimation algorithms to this case. While estimating common factors remains completely unchanged, the QTT estimator only requires a minor modification of \eqref{QTEstep2}, which becomes:
\begin{align} \label{QTEstep2-extension1}
    (\widecheck{\lambda}_{1},\widecheck{\zeta}) = \underset{(\lambda_1, \zeta)}{\operatorname{argmin}} \sum_{t=1}^T \rho_{\tau}(y_{1t} - \lambda_1'\widecheck{f}_{t} - g(t;\zeta) d_{1t}),
\end{align}
where $\delta_t=g(t;\zeta)$ is parametrized by $\zeta$. However, including a function of $t$ in the quantile regression may lead to a different asymptotic property of the estimator $(\widecheck{\lambda}_{1},\widecheck{\zeta})$, which requires more involved and case-by-case theoretical analysis. For example, for $g(t)=\zeta_0+\zeta_1t$, the convergence rate for $\widehat{\zeta}_1$ would be $T^{3/2}$ instead of $T^{1/2}$ \citep{hamilton2020time}.

\subsubsection{QTT as A Function of Factors}\label{sec:factor-variant}

We could also allow QTT to be a function of common factors. For example, we may extend \eqref{qfs_linear} to be $Q_{\tau}(y^1_{it}|f_t) =\zeta_0 + (\lambda_i+\zeta_1)' f_t$. In other words, the treatment may influence the potential outcome through the factor loadings as well, and the true QTT becomes $\delta_t=\zeta_0+\zeta_1'f_t$. This is analogous to adding an interactive term in the linear model. For estimation, we can still apply Algorithm \ref{algo-QTE}, with the following modification in Step 2:
    \begin{align} \label{eq:extension-step2}
        (\widecheck{\lambda}_{1},\widecheck{\zeta}_0, \widecheck{\zeta}_1) = \underset{(\lambda_1,\zeta_0,\zeta_1)}{\operatorname{argmin}} \sum_{t=1}^T \rho_{\tau}(y_{1t}-\lambda_1'\widecheck{f}_{t} -\zeta_1'\widecheck{f}_{t} d_{1t} - \zeta_0 d_{1t}).
    \end{align}
One may consider more complicated interactions between treatment and factors, or even semi/nonparametric functional forms, and our theoretical results can be extended to this case. However, caution should be taken when interpreting the parameters $\zeta$, since $f_t$ are unobserved latent factors and lack real-world meaning.

\subsection{Multiple Treated Units}
\label{sec:multiple}

Our QTT estimation method could be easily extended to the multiple treated units case, which typically raises heavy computational burdens in SCM. Under the traditional weighting scheme SCM, the weights might be non-unique when there are multiple treated units \citep{abadie2021using}. 
To address the issue of multiplicity, \cite{abadie2021penalized} propose the use of penalized regression.
Such penalized regressions should be executed for every treated unit $i$ that may bring in substantial computation as in \cite{cai2022estimating}. 
In contrast, our method uses the factors $f_t$ capture the common underlying characteristics in all units, which is estimated only once using the control group. In particular, to extend our method to the case with multiple treated units, we just need to apply Algorithm \ref{algo-QTE} for every treated unit $i$:
\begin{align} 
    (\widecheck{\lambda}_{i}(\tau),\widecheck{\delta}_{i}(\tau)) = \underset{(\lambda_i,\delta_{i})}{\argmin} \sum_{t=1}^T \rho_{\tau}(y_{it}-\lambda_i'\widecheck{f}_{t}(\tau)-\delta_{i} d_{it}).
\end{align}
Since the computation of quantile regression is relatively efficient, our QTT estimators could be easily adapted to the case with multiple treated units.

\subsection{Observed Covariates}
\label{sec:covariates}
While our method does not require predictors for $y_{it}$, incorporating information from observed covariates might assist the estimation of the quantile factor model. Suppose we have a $K$-dimensional vector of continuous covariates $x_t=(x_{1t},...,x_{Kt})'$ which shares the same quantile factor structure with $y_{it}^0$:
\begin{align}
\label{eqn:covariates}
    Q_{\tau}(x_{kt}|f_t(\tau)) = \mu_k(\tau)'f_t(\tau), k=1,...,K.
\end{align}
Then our method could be directly extended by adding $x_t$ to the control units $(y_{it})_{i=2}^{N+1}$ to estimate the latent factors $f_t(\tau)$. Good candidates of $x_t$ may include predictors for the outcome of the treated unit $y_{1t}$, or variables that capture the common time trends. 
As always, substantive knowledge is helpful for finding good covariates. For instance, in the asset pricing literature, some well-known factors, such as those proposed by Fama and French, are found to be significantly correlated with the factors estimated from empirical studies \citep{ando2020quantile}.

\cite{ando2020quantile} consider a quantile regression model with interactive fixed effects which is close to the quantile factor model of \cite{chen2021quantile} and allow for covariates. Since our goal is QTT, recovering the quantile factor structure from the control units using \cite{chen2021quantile}'s quantile factor model is sufficient for us to uncover the latent factor structure for the treated unit. Moreover, \cite{chen2021quantile}'s method has the advantage of efficiently estimating for the number of factors and allowing for heavy tails in error terms. Nonetheless, including covariates using the framework of \cite{ando2020quantile} is a promising research revenue and is left for future exploration.

\section{Conclusions}
\label{sec:conclusion}
This paper proposes quantile treatment effects on treated (QTT) estimators under high dimensional panel data, which extends \cite{xu_2017}'s generalized synthetic control method (GSCM) for estimating average treatment effects on treated (ATT). Our framework estimates QTT in a panel data setting where $N,T\to\infty$ and allows the underlying common factor structure to change with quantiles, where the issue of high-dimensionality is handled via the first-stage estimation of a quantile factor model. We provide asymptotic properties for our proposed estimators, along with a valid inference approach based on blockwise bootstrap. Monte-Carlo simulations demonstrate the effectiveness of our method, which is suitable for studying policy effect in the fields such as macroeconomics and finance where high frequency panel data can be approached. We apply our method to study the impact of the 2008 China Stimulus Program's on China's economy. Our empirical results show that QTT serves as a valuable complement to ATT.

However, our proposal exhibits certain limitations. First, estimations in small sample sizes at the tail quantiles may be less precise, as shown in the simulations. Second, There is potential for improvement in the blockwise bootstrap method, particularly in cases of small samples where the inference at some quantile levels are not precise enough. 
Moreover, for the QTT estimator $\widehat{\delta}$ based on Algorithm \ref{algo-IQR}, we only provide a consistency result due to technical reasons illustrated in Section \ref{sec:framework}. We conjecture that asymptotic normality may hold under certain conditions and we show its valid inference via simulations, while a formal proof is highly encouraged in the future study.

Beyond the extensions discussed in Section \ref{sec:extension}, we believe that the core idea developed in this paper can be fruitfully applied to other emerging panel data causal inference literature, particularly those based on low-rank factor structures, such as matrix completion \citep{athey2021matrix}. However, such extensions might not be straightforward. While matrix completion determines the rank adaptively through regularization and leverages information from both treated and control units in estimating the factor structure, its implementation requires iterative procedures (see Section 4.2 in \citealp{athey2021matrix}) involving matrix shrinkage operators, which may become computationally burdensome when combined with the IQR/ISQR framework. Furthermore, inference in matrix completion typically relies on subsampling methods and lacks a formal asymptotic theory. These challenges highlight that extending these new-generation panel causal estimators into factor-based QTT frameworks is a nontrivial yet promising direction for future research.

\bibliography{QTE-FM-ref}

\newpage
\appendixpage

\appendix
\section{Proofs} \label{A-proof}

We first introduce some useful lemmas for our proofs. Recall that $\widetilde{f}_t$ is estimated from Algorithm \ref{algo-ISQR}.

\begin{lemma}
\label{lemma1}
    Under Assumptions \ref{asmp-factor} \ref{asmp-density}, \ref{asmp-error}, \ref{asmp-smooth}, we have $T^{-1}\sum_{t=1}^T c_t(\widetilde{f}_t-f_{0t})=O_p(T^{-1}h^{-1})$ for uniformly bounded constants $\{c_t\}_{t=1}^T$.
\end{lemma}

\begin{proof}[Proof]
    See Lemma S.7 in the Supplement of \cite{chen2021quantile}. 
\end{proof}

\begin{lemma}
\label{lemma3}
    Under Assumptions \ref{asmp-factor},  \ref{asmp-density}, \ref{asmp-error}, \ref{asmp-smooth}, $\|\widetilde{f}_t-f_{0t}\|=O_p(T^{-1/2}h^{-1})$ for each $t$.
\end{lemma}
\begin{proof}
    See Lemma S.5 in Supplement of \cite{chen2021quantile}.
\end{proof}

Now we present the proofs of our main theorems. 
\begin{proof}[Proof of Theorem \ref{thm1}.]
    Let $M_T(\beta)=T^{-1}\sum_{t=1}^T \rho_\tau(y_{1t}-\widehat{w}_t'\beta)$, where $\widehat{w}_t=(\widehat{f}_t',d_{1t})'$. Let  $\overline{M}_T(\beta)=T^{-1}\sum_{t=1}^T\rho_\tau(y_{1t}-(\widehat{S}_0\widehat{w}_t)'\beta)$, and $M(\beta)=\mathbb{E}[\rho_\tau(y_{1t}-w_{0t}'\beta)]$. Denote $\mathcal{B}=\mathcal{A}\times\Delta$ as the parameter space for $\beta$. Suppose $\overline{\beta}=\argmin_{\beta\in \mathcal{B}}\overline{M}_T(\beta)$ and recall that $\widehat{\beta}=\argmin_{\beta\in \mathcal{B}}M_T(\beta)$, then $\widehat{\beta}=\widehat{S}_0'\overline{\beta}$. Since $\widehat{S}_0$ is a diagonal matrix with the last element being $1$, then $\widehat{\delta}=\overline{\delta}$ where $\widehat{\delta}$ is the last element of $\widehat{\beta}$ and $\overline{\delta}$ is the last element of $\overline{\beta}$. Thus, it suffices to show that $\overline{\beta}\stackrel{p}{\to}\beta_0$ where $\beta_0=\argmin_{\beta\in \mathcal{B}}M(\beta)$ and $\delta_0$ is the last element of $\beta_0$. 
    Therefore, by Theorem 5.7 of \cite{van2000asymptotic} we only need to show
    \begin{align}
        \sup_{\beta\in \mathcal{B}} |\overline{M}_T(\beta)-M(\beta)|\stackrel{p}{\to}1.
    \end{align}
    Note that
    \begin{align}
        \nonumber |\overline{M}_T(\beta)-M(\beta)| 
        &= \bigg| \frac{1}{T}\sum_{t=1}^T \big\{ \rho_{\tau}(y_{1t}-(\widehat{S}_0\widehat{w}_t)'\beta)-\mathbb{E}[\rho_\tau(y_{1t}-w_{0t}'\beta)]\big\} \bigg| \\
        \nonumber &\leq \bigg|\frac{1}{T}\sum_{t=1}^T \big\{\rho_{\tau}(y_{1t}-(\widehat{S}_0\widehat{w}_t)'\beta)-\rho_{\tau}(y_{1t}-w_{0t}'\beta)\big\}\bigg| \\
        \nonumber &+ \bigg|\frac{1}{T}\sum_{t=1}^T\big\{ \rho_{\tau}(y_{1t}-w_{0t}'\beta)-\mathbb{E}[\rho_\tau(y_{1t}-w_{0t}'\beta)]\big\} \bigg| \\
        &\equiv |R_{1}(\beta)| + |R_2(\beta)|.
    \end{align}
    From Lemma 1(ii) of \cite{gregory2018smooth}, $\sup_{\beta\in \mathcal{B}}|R_2(\beta)|\stackrel{p}{\to}1$ using Assumption \ref{asmp-weakdep} and \ref{asmp-iden}. Next, using Knight's identity \citep{knight1998limiting}:
    \begin{align}
        \rho_\tau(u-v)-\rho_\tau(u) = -v\psi_\tau(u) + \int_0^v [\mathbf{1}(u\leq s)-\mathbf{1}(u\leq 0)]ds,
    \end{align}
    where $\psi(u)=\tau-\mathbf{1}(u\leq 0)$. Let $u=y_{1t}-w_{0t}'\beta$ and $v=[\widehat{S}_0(\widehat{w}_t-(\widehat{S}_0)^{-1}w_{0t})]'\beta$. Noticing that $\widehat{S}_0=\widehat{S}_0^{-1}$ and $\widehat{S}_0=\widehat{S}_0'$, we have $v=(\widehat{w}_t-\widehat{S}_0w_{0t})'\widehat{S}_0\beta$. Hence,
    \begin{align}
        \nonumber |R_1(\beta)| &= \bigg|-\frac{1}{T}\sum_{t=1}^T (\widehat{w}_t-\widehat{S}_0w_{0t})'\widehat{S}_0\beta\psi_\tau(y_{1t}-w_{0t}'\beta) \\
        \nonumber &+ \frac{1}{T}\sum_{t=1}^T\int_{0}^{(\widehat{w}_t-\widehat{S}_0w_{0t})'\widehat{S}_0\beta}[\mathbf{1}(y_{1t}-w_{0t}'\beta\leq s)-\mathbf{1}(y_{1t}-w_{0t}'\beta\leq 0)]ds \bigg| \\
        \nonumber &\leq \bigg|\frac{2}{T}\sum_{t=1}^T (\widehat{w}_t-\widehat{S}_0w_{0t})'\widehat{S}_0\beta\bigg| + \bigg|\frac{2}{T}\sum_{t=1}^T (\widehat{w}_t-\widehat{S}_0w_{0t})'\widehat{S}_0\beta \bigg| \\
        \nonumber &\leq 4\|\beta\|\|\widehat{S}_0\|\bigg(\frac{1}{T}\sum_{t=1}^T\|\widehat{w}_t-\widehat{S}_0w_{0t}\|^2\bigg)^{1/2} \\
        &= O_p(L_{NT}^{-1})=o_p(1),
    \end{align}
    where $L_{NT}=\min\{\sqrt{N},\sqrt{T}\}$, and the last line follows from Theorem 1 of \cite{chen2021quantile} and Assumptions \ref{asmp-factor}-\ref{asmp-error}. This completes the proof.
\end{proof}

\begin{proof}[Proof of Theorem \ref{thm2}.]
Consider estimating the common factor using Algorithm \ref{algo-ISQR} and obtain $\widetilde{f}_t$, and let $\widetilde{w}_t=(\widetilde{f}_t',d_{1t})'$. The QTT estimation in Algorithm \ref{algo-QTE} leads to the optimization problem
\begin{align}
    \widetilde{\beta} = \underset{\beta\in\mathbbm{R}^{r+1}}{\text{argmin}}  \sum_{t=1}^T \rho_{\tau}(y_{1t}-\widetilde{w}_t'\beta).
\end{align}
Define
\begin{align}
    L(\gamma) = \sum_{t=1}^T \rho_\tau (\epsilon_{1t} - \widetilde{w}_t'\gamma/\sqrt{T}-v_t)-\rho_{\tau}(\epsilon_{1t}),
\end{align}
where $v_t=(\widetilde{w}_t-w_{0t})'\beta_0$ and $\beta_0=(\lambda_{01}',\delta_0')'$. Then $\widetilde{\gamma}\equiv\sqrt{T}(\widetilde{\beta}-\beta_0)$ is the unique minimizer of $L(\gamma)$. By the argmax theorem \citep[Theorem 5.56]{van2000asymptotic}, if $L(\gamma)\stackrel{d}{\to}L_0(\gamma)$, and $\gamma_0$ is the unique minimizer of $L_0(\gamma)$, then $\widetilde{\gamma}\stackrel{d}{\to}\gamma_0$. By Knight's identity, we have
\begin{align}
    \nonumber L(\gamma) &= \sum_{t=1}^T -(\widetilde{w}_t'\gamma/\sqrt{T}+v_t) \psi_\tau(\epsilon_{1t}) \\
    \nonumber &+ \sum_{t=1}^T \int_0^{\widetilde{w}_t'\gamma/\sqrt{T}+v_t}[\mathbf{1}(\epsilon_{1t}\leq s)-\mathbf{1}(\epsilon_{1t}\leq 0)]ds \\
    &\equiv L_1(\gamma) + L_2(\gamma).
\end{align}
First consider $L_1(\gamma)$. Since $v_t$ is irrelevant of $\gamma$,
\begin{align}
    \nonumber L_1(\gamma) &= \sum_{t=1}^T -(\widetilde{w}_t'\gamma/\sqrt{T}+v_t) \psi_\tau(\epsilon_{1t}) \\
    \nonumber &= -\frac{1}{\sqrt{T}}\sum_{t=1}^T \widetilde{w}_t'\gamma \psi_\tau(\epsilon_{1t}) - \sum_{t=1}^T v_t\psi_{\tau}(\epsilon_{1t}) \\
    \nonumber &= -\frac{1}{\sqrt{T}}\sum_{t=1}^T w_{0t}'\gamma \psi_\tau(\epsilon_{1t}) - \frac{1}{\sqrt{T}}\sum_{t=1}^T (\widetilde{w}_t-w_{0t})'\gamma \psi_\tau(\epsilon_{1t}) - \sum_{t=1}^T v_t\psi_{\tau}(\epsilon_{1t}) \\
    &\equiv L_{11}(\gamma) + L_{12}(\gamma) + L_{13}.
\end{align}
By Lemma \ref{lemma1} and Assumption \ref{asmp-smooth}(v), $L_{12}(\gamma)=O_p(T^{-1/2}h^{-1})\|\gamma\|=o_p(T^{-1/2+1/6})\|\gamma\|=o_p(\|\gamma\|)$. 
For $L_{11}(\gamma)$, we first observe that by Assumption \ref{asmp-iden}(i) and the quantile condition $Q_\tau(\epsilon_{1t}|f_{0t})=0$, we have $\mathbb{E}[\psi_\tau(\epsilon_{1t})|f_{0t},d_{1t}]=\mathbb{E}[\psi_\tau(\epsilon_{1t})|f_{0t}]=0$. 
Since strict stationarity and $\alpha$-mixing is preserved under any measurable transformation \citep[Theorem 15.1]{davidson1994stochastic}, along with Assumption \ref{asmp-weakdep}, we have $\psi_\tau(\epsilon_{1t})$ is strictly stationary and $\alpha$-mixing with the same mixing coefficients. Since $\mathbb{E}[|\psi_\tau(\epsilon_{1t})|^s|w_{0t}]\leq 2$ for any $s\geq 1$ and $\sum_{\ell=1}^{\infty}\alpha(\ell)^{1-2s}\leq \sum_{\ell=1}^{\infty}\alpha(\ell)^{1-\kappa}<\infty$ for some large $s$ by Assumption \ref{asmp-weakdep}, the conditions in central limit theorem for strong mixing process \citep[Theorem 14.15]{hansen2022econometrics} is satisfied. Along with Assumption \ref{asmp-var}, we have
\begin{align}
\label{eqn:L11}
    L_{11}(\gamma) \stackrel{d}{\to} -\gamma'W,
\end{align}
where $W\sim \mathcal{N}(0,\Sigma)$. Next, consider $L_2(\gamma)$. Since $\widetilde{\beta}-\beta_0=o_p(1)$ which follows a similar argument as in Theorem \ref{thm1}, and that $v_t=O_p(T^{-1}h^{-1})=o_p(1)$ by Lemma \ref{lemma1} and Assumption \ref{asmp-smooth}(v), we have $\widetilde{w}_t'\gamma/\sqrt{T}+v_t=o_p(1)$ at $\gamma=\widetilde{\gamma}$. Using Taylor expansion and Assumption \ref{asmp-density},
\begin{align}
\label{eq:L2}
    \nonumber \mathbb{E}[L_2(\gamma)|F_0] &= \sum_{t=1}^T \int_0^{\widetilde{w}_t'\gamma/\sqrt{T}+v_t}[F_{\epsilon_{1t}}(s|f_{0t})-F_{\epsilon_{1t}}(0|f_{0t})]ds \\
    \nonumber &= \sum_{t=1}^T \int_0^{\widetilde{w}_t'\gamma/\sqrt{T}+v_t}[h_{1t}(0|f_{0t})s + O(s^2)]ds \\
    &= \sum_{t=1}^T \bigg[\frac{1}{2}h_{1t}(0|f_{0t})(\widetilde{w}_t'\gamma/\sqrt{T}+v_t)^2 + O((\widetilde{w}_t'\gamma/\sqrt{T}+v_t)^3) \bigg],
\end{align}
Since $\widetilde{w}_t'\gamma/\sqrt{T}+v_t=o_p(1)$, the second term in \eqref{eq:L2} is of smaller order of first term. By Lemma \ref{lemma3}, we have $\max_t\|\widetilde{w}_t\|\leq \max_t \|w_{0t}\| + \max_t \|\widetilde{w}_t-w_{0t}\|=O_p(1)$, leading to the cross term in the first term being
\begin{align}
    \nonumber \frac{1}{\sqrt{T}}\sum_{t=1}^T h_{1t}(0|f_{0t})v_t\widetilde{w}_t'\gamma &= \frac{1}{\sqrt{T}}\sum_{t=1}^T (\widetilde{w}_t-w_{0t})'\beta_0h_{1t}(0|f_{0t})\widetilde{w}_t'\gamma \\
    \nonumber &\leq \|\gamma\| \max_{t\in\{1,...,T\}}\|\widetilde{w}_t\| \frac{1}{\sqrt{T}}\sum_{t=1}^T (\widetilde{w}_t-w_{0t})'\beta_0h_{1t}(0|f_{0t}) \\
    &= \|\gamma\| O_p(1) O_p(T^{-1/2}h^{-1}) =o_p(\|\gamma\|),
\end{align}
where the last line comes from Lemma \ref{lemma1} and Assumption \ref{asmp-smooth}(v).
Now consider
\begin{align}
    \nonumber Z(\gamma) &= \frac{1}{2} \sum_{t=1}^T \bigg[h_{1t}(0)(\widetilde{w}_t'\gamma/\sqrt{T})^2\bigg] \\
    \nonumber & = \gamma'\bigg[\frac{1}{2T}\sum_{t=1}^T h_{1t}(0) \widetilde{w}_t\widetilde{w}_t' \bigg]\gamma \\
    \nonumber &= \gamma'\bigg[\frac{1}{2T}\sum_{t=1}^T h_{1t}(0) w_{0t}w_{0t}' \bigg]\gamma + \gamma'\bigg[\frac{1}{2T}\sum_{t=1}^T h_{1t}(0) (\widetilde{w}_t-w_{0t})(\widetilde{w}_t-w_{0t})' \bigg]\gamma \\
    \nonumber &+ \gamma'\bigg[\frac{1}{2T}\sum_{t=1}^T h_{1t}(0) (\widetilde{w}_t-w_{0t})w_{0t}' \bigg]\gamma \\
    &\equiv Z_1(\gamma) + Z_2(\gamma) + Z_3(\gamma).
\end{align}
By Assumption \ref{asmp-iden}(ii), $Z_1(\delta)\to\gamma'\Gamma\gamma/2$. By Lemma \ref{lemma1} and Assumption \ref{asmp-smooth}(v), 
\begin{align}
    \nonumber Z_2(\gamma) &\leq \frac{1}{2}\overline{h}\|\gamma\|^2 \bigg\|\frac{1}{T}\sum_{t=1}^T (\widetilde{w}_t-w_{0t})(\widetilde{w}_t-w_{0t})'\bigg\| \\
    \nonumber &= \frac{1}{2}\overline{h}\|\gamma\|^2 \frac{1}{T}\sum_{t=1}^T \|\widetilde{w}_t-w_{0t}\|^2 \\
    &=\|\gamma\|^2O_p(T^{-1}h^{-1})=o_p(\|\gamma\|^2).
\end{align}
Similarly,
\begin{align}
    \nonumber Z_3(\gamma) &\leq \frac{1}{2}\overline{h}\|\gamma\|^2\frac{1}{T}\sum_{t=1}^T \|\widetilde{w}_t-w_0\|\|w_{0t}\| \\
    &= \|\gamma\|^2 O_p(T^{-1/2}h^{-1})=o_p(\|\gamma\|^2)
\end{align}
by Lemma \ref{lemma3}.
Thus, combining the above results, we have
\begin{align}
    L(\gamma) \stackrel{d}{\to} L_0(\gamma) =-\gamma'W + \frac{1}{2}\gamma'\Gamma\gamma,
\end{align}
with $\gamma_0=\Gamma^{-1}W$ being the unique minimizer of $L_0(\gamma)$ by its convexity \citep{koenker2005quantile}. This completes the proof.
\end{proof}

\begin{proof}[Proof of Theorem \ref{thm3}]
We proceed by the following steps.

    \noindent\textbf{Step 1.} Let $\vec{\beta}=\argmin_{\beta\in \mathcal{B}}\sum_{t=1}^T\rho_\tau(y_{1t}-w_{0t}'\beta)$, and $\vec{\beta}^\ast$ be the block bootstrap version of $\vec{\beta}$ which uses the same block positions as $\widetilde{\beta}^\ast$. Then $\vec{\beta}^\ast$ is the unsmoothed block bootstrap, where the bootstrap centering is just $\vec{\beta}$ \citep[p. 1147]{gregory2018smooth}. Here we perform block bootstrap on each value of the non-random variable $d_{1t}=\{0,1\}$ so that we have approximately the same treated and control units in the bootstrap sample, while the bootstrap sample still mimics the population. Under Assumptions \ref{asmp-factor}-\ref{asmp-boot}, using Theorem 2 of \cite{gregory2018smooth}, we have
    \begin{align}
        \sup_{z\in\mathbb{R}^{r+1}}\bigg|\mathbb{P}^\ast\big[\sqrt{T}(\vec{\beta}^\ast-\vec{\beta})\leq z\big] - \mathbb{P}\big[\sqrt{T}(\vec{\beta}-\beta_0)\leq z\big]\bigg|\stackrel{p}{\to} 0,        
    \end{align}

    \noindent\textbf{Step 2.} We show that $\sqrt{T}(\widetilde{\beta}-\vec{\beta})=o_p(1)$. From the proof of Theorem \ref{thm2}, we have
    \begin{align}
        L(\gamma) = \gamma'\bigg[-\frac{1}{\sqrt{T}}\sum_{t=1}^T w_{0t}\psi_\tau(\epsilon_{1t})\bigg] + o_p(\|\gamma\|) + \gamma'\bigg[\frac{1}{2T}\sum_{t=1}^T h_{1t}(0) w_{0t}w_{0t}' \bigg]\gamma + o_p(\|\gamma\|^2).
    \end{align}
    Therefore, the first order condition yields
    \begin{align}
        \sqrt{T}(\widetilde{\beta}-\beta_0) = \Gamma^{-1} \frac{1}{\sqrt{T}}\sum_{t=1}^T w_{0t}\psi_\tau(\epsilon_{1t}) + o_p(1).
    \end{align}
    On the other hand, the standard quantile regression also results in the same Bahadur representation \citep{koenker1978regression}:
    \begin{align}
        \sqrt{T}(\vec{\beta}-\beta_0) = \Gamma^{-1} \frac{1}{\sqrt{T}}\sum_{t=1}^T w_{0t}\psi_\tau(\epsilon_{1t}) + o_p(1).
    \end{align}
    The result follows.
    
    \noindent\textbf{Step 3.} We show that $\sqrt{T}(\widetilde{\beta}^\ast-\vec{\beta}^\ast)=o_p^\ast(1)$. Since $\widetilde{\beta}^\ast$ and $\vec{\beta}^\ast$ are generated from the same random block positions, where the former uses $\widetilde{f}^\ast_t$ and the latter uses $f^\ast_{0t}$ as regressors, from Step 2, we have $\sqrt{T}(\widetilde{\beta}^\ast-\vec{\beta}^\ast)=o_p(1)$ for any given block positions. Thus, taking into account the randomness from the bootstrap, we still have $\sqrt{T}(\widetilde{\beta}^\ast-\vec{\beta}^\ast)=o_p^\ast(1)$ in probability, that is, $\sqrt{T}(\widetilde{\beta}^\ast-\vec{\beta}^\ast)=r_{1T}^\ast$ where $\mathbb{P}^\ast\big[|r^\ast_{1T}|>z\big]\stackrel{p}{\to}0$ for any $z>0$.

    \noindent\textbf{Step 4.} We show that
    \begin{align}
        \sup_{z\in\mathbb{R}^{r+1}}\bigg|\mathbb{P}^\ast\big[\sqrt{T}(\widetilde{\beta}^\ast-\widetilde{\beta})\leq z\big] - \mathbb{P}^\ast\big[\sqrt{T}(\vec{\beta}^\ast-\vec{\beta})\leq z\big]\bigg|\stackrel{p}{\to} 0.
    \end{align}
    From Step 2 and 3, we have $\sqrt{T}(\widetilde{\beta}^\ast-\widetilde{\beta})=\sqrt{T}(\vec{\beta}^\ast-\vec{\beta})+r^\ast_{1T}+r_{2T}$. Let $r^\ast_T=r_{1T}^\ast+r_{2T}$, then
    \begin{align}
        \nonumber \mathbb{P}^\ast\big[\sqrt{T}(\widetilde{\beta}^\ast-\widetilde{\beta})\leq z\big] &= \mathbb{P}^\ast\big[\sqrt{T}(\vec{\beta}^\ast-\vec{\beta})+r^\ast_T\leq z\big] \\
        \nonumber & \leq \mathbb{P}^\ast\big[\sqrt{T}(\vec{\beta}^\ast-\vec{\beta}) - |r^\ast_T| \leq z\big] \\
        & \leq \mathbb{P}^\ast\big[\sqrt{T}(\vec{\beta}^\ast-\vec{\beta}) \leq z\big] + \mathbb{P}^\ast\big[|r^\ast_T|>z\big].
    \end{align}
    On the other hand, 
    \begin{align}
        \nonumber \mathbb{P}^\ast\big[\sqrt{T}(\widetilde{\beta}^\ast-\widetilde{\beta})\leq z\big] & \geq \mathbb{P}^\ast\big[\sqrt{T}(\vec{\beta}^\ast-\vec{\beta}) + |r^\ast_T| \leq z\big] \\
        \nonumber & \geq \mathbb{P}^\ast\big[\sqrt{T}(\vec{\beta}^\ast-\vec{\beta}) \leq z\big] - \mathbb{P}\big[|r^\ast_T|>z\big].
    \end{align}
    Since $\mathbb{P}\big[|r^\ast_T|>z\big]\stackrel{p}{\to}0$ as $T\to\infty$, the result follows.
    
    \noindent\textbf{Step 5.} From Theorem \ref{thm2}, $\sqrt{T}(\widetilde{\beta}-\beta_0)$ and $\sqrt{T}(\vec{\beta}-\beta_0)$ have the same asymptotic distribution. Thus, we have
    \begin{align}
        \sup_{z\in\mathbb{R}^{r+1}}\bigg|\mathbb{P}\big[\sqrt{T}(\vec{\beta}-{\beta}_0)\leq z\big] - \mathbb{P}\big[\sqrt{T}(\widetilde{\beta}-\beta_0)\leq z\big]\bigg|\stackrel{p}{\to}0.
    \end{align}
    Combining the above results, we have
    \begin{align}
        \nonumber &         \sup_{z\in\mathbb{R}^{r+1}}\bigg|\mathbb{P}^\ast\big[\sqrt{T}(\widetilde{\beta}^\ast-\widetilde{\beta})\leq z\big] - \mathbb{P}\big[\sqrt{T}(\widetilde{\beta}-\beta_0)\leq z\big]\bigg| \\
        \nonumber &\leq \sup_{z\in\mathbb{R}^{r+1}}\bigg|\mathbb{P}^\ast\big[\sqrt{T}(\widetilde{\beta}^\ast-\widetilde{\beta})\leq z\big] - \mathbb{P}^\ast\big[\sqrt{T}(\vec{\beta}^\ast-\vec{\beta})\leq z\big]\bigg| \\
        \nonumber &+ \sup_{z\in\mathbb{R}^{r+1}}\bigg|\mathbb{P}^\ast\big[\sqrt{T}(\vec{\beta}^\ast-\vec{\beta})\leq z\big] - \mathbb{P}\big[\sqrt{T}(\vec{\beta}-\beta_0)\leq z\big]\bigg| +  \\
        \nonumber &+ \sup_{z\in\mathbb{R}^{r+1}}\bigg|\mathbb{P}\big[\sqrt{T}(\vec{\beta}-\beta_0)\leq z\big] - \mathbb{P}\big[\sqrt{T}(\widetilde{\beta}-\beta_0)\leq z\big]\bigg| \\
        &=o_p(1)
    \end{align}
    
\end{proof}

\section{Additional Simulations Results}

In this section, we report some further Monte Carlo simulations regarding the performance of \emph{NQTT} and \emph{SQTT} for the following cases: (1) the idiosyncratic error exhibits heavy tails; (2) the idiosyncratic errors are serially correlated; and (3) we have a more complicated quantile variant factor structure.

\subsection{Errors with Heavy Tails} \label{sec:t-tail}

We preserve the DGP setup in Section \ref{sec:simulation} while changing the error term to be $u_{it}\stackrel{\text{i.i.d}}{\sim}t(2)$, which is a student-t distribution with two degrees of freedom. As illustrated in Section \ref{sec:asymptotic}, our estimation approach does not require moments of error terms to exist, thus allowing for heavy-tailed idiosyncratic shocks.

Table \ref{tb1} delivers similar conclusions as Table \ref{tab-1} that, at the tail quantiles, our estimators largely outperforms \emph{GSCM}, and at the middle quantile $\tau=0.5$ both the QTTs and \emph{GSCM} work satisfactorily. In addition, \emph{NQTT} and \emph{SQTT} are close to \emph{Oracle} at all quantiles, verifying the effectiveness of our proposed method under heavy-tailed error distributions. Table \ref{tb2} indicates that the bootstrap method valid as in Table \ref{tab-2}.

\begin{table}[!t] 
\centering
\caption{Bias and RMSE: $u_{it}\stackrel{\text{i.i.d}}{\sim}t(2)$}
\label{tb1}
\begin{tabular}{cccccccccc}
  \toprule
   \multicolumn{3}{c}{$\tau$} & 0.1 & 0.25 & 0.5 & 0.75 & 0.9  \\ 
   
  \midrule

\multirow{8}{*}{\makecell{N=50 \\ T=100}} 
& \multirow{2}{*}{NQTT} & Bias & 0.0423 & -0.0816 & -0.0101 & 0.0936 & -0.0755  \\
& & RMSE &1.1274 & 0.5440 & 0.3652 & 0.5813 & 1.2209   \\ \cmidrule{2-8}

& \multirow{2}{*}{SQTT} & Bias & 0.0701 & -0.0941 & -0.0073 & 0.1052 & -0.0847  \\
& & RMSE &1.1475 & 0.5572 & 0.3676 & 0.5719 & 1.2027  \\ \cmidrule{2-8}

& \multirow{2}{*}{Oracle} & Bias &0.0017 & 0.0079 & -0.0039 & 0.0029 & 0.0149     \\
& & RMSE & 1.0346 & 0.5038 & 0.3939 & 0.5319 & 1.0884   \\ 
\cmidrule{2-8}

& \multirow{2}{*}{GSCM}  & Bias &-1.9383 & -0.6774 & 0.0022 & 0.6866 & 1.9663   \\
& & RMSE &2.3490 & 0.8707 & 0.3703 & 0.8978 & 2.3712   \\
\midrule

\multirow{8}{*}{\makecell{N=100 \\ T=200}} 
& \multirow{2}{*}{NQTT} & Bias &0.2242 & -0.0598 & -0.0003 & 0.0820 & -0.2024    \\
& & RMSE &0.9538 & 0.4464 & 0.3019 & 0.4347 & 0.9594    \\ \cmidrule{2-8}

& \multirow{2}{*}{SQTT} & Bias & 0.1683 & -0.0616 & 0.0007 & 0.0919 & -0.1306  \\
& & RMSE &1.0706 & 0.4436 & 0.3036 & 0.4369 & 1.0736 \\ \cmidrule{2-8}

& \multirow{2}{*}{Oracle} & Bias & 0.0007 & 0.0122 & 0.0121 & 0.0341 & 0.0472    \\
& & RMSE &0.8705 & 0.4210 & 0.3131 & 0.4246 & 0.9027   \\ \cmidrule{2-8}

& \multirow{2}{*}{GSCM} & Bias &-3.0518 & -1.1269 & -0.0019 & 1.1214 & 3.0152   \\
& & RMSE &3.2696 & 1.2386 & 0.3372 & 1.2343 & 3.2305    \\
\midrule

\multirow{8}{*}{\makecell{N=200 \\ T=400}} 

& \multirow{2}{*}{NQTT} & Bias & 0.0712 & -0.0169 & -0.0015 & 0.0279 & -0.0502  \\
& & RMSE &0.6195 & 0.2601 & 0.1871 & 0.2585 & 0.6230   \\ \cmidrule{2-8}

& \multirow{2}{*}{SQTT} & Bias & 0.1392 & -0.0227 & -0.0021 & 0.0412 & -0.1143  \\
& & RMSE &0.7088 & 0.3114 & 0.2216 & 0.3146 & 0.7625 \\ \cmidrule{2-8}

& \multirow{2}{*}{Oracle}  & Bias   & 0.0109 & 0.0012 & 0.0058 & 0.0101 & 0.0161   \\
& & RMSE  &0.4923 & 0.2527 & 0.2010 & 0.2592 & 0.5518    \\
\cmidrule{2-8}

& \multirow{2}{*}{GSCM}  & Bias    & -2.0089 & -0.7482 & -0.0010 & 0.7395 & 2.0091 \\
& & RMSE &  2.1002 & 0.7955 & 0.1912 & 0.7913 & 2.1140  \\
\bottomrule

\end{tabular}
\end{table}

\begin{table}[!t] 
\centering
\caption{Standard Deviation and 95\% Coverage for Bootstrap: $u_{it}\stackrel{\text{i.i.d}}{\sim}t(2)$}
\label{tb2}
\begin{tabular}{ccccccccccc}
  \toprule
&   \multicolumn{2}{c}{$\tau$} & 0.1 & 0.25 & 0.5 & 0.75 & 0.9  \\ 
   
  \midrule

\multirow{4}{*}{\makecell{N=50 \\ T=100}}  

& \multirow{2}{*}{NQTT} & SD   &1.6220 & 0.5765 & 0.3836 & 0.5849 & 1.6250  \\   
& & Coverage & 0.9480 & 0.9500 & 0.9460 & 0.9540 & 0.9340 \\   
\cmidrule{2-8}

& \multirow{2}{*}{SQTT} & SD   &1.6215 & 0.5767 & 0.3844 & 0.5845 & 1.5867\\   
& & Coverage &0.9420 & 0.9510 & 0.9480 & 0.9540 & 0.9430 \\   
\midrule 

\multirow{4}{*}{\makecell{N=100 \\ T=200}}  

& \multirow{2}{*}{NQTT} & SD   & 1.0367 & 0.4616 & 0.3076 & 0.4654 & 1.0463 \\    
& & Coverage &0.9330 & 0.9590 & 0.9510 & 0.9570 & 0.9380 \\ \cmidrule{2-8}

& \multirow{2}{*}{SQTT} & SD   &1.0307 & 0.4613 & 0.3063 & 0.4650 & 1.0365\\   
&  & Coverage &0.9280 & 0.9610 & 0.9530 & 0.9580 & 0.9190  \\   
\midrule

\multirow{4}{*}{\makecell{N=200 \\ T=400}}  

& \multirow{2}{*}{NQTT} & SD   &0.6075 & 0.2676 & 0.1854 & 0.2620 & 0.6164   \\
& & Coverage &0.9340 & 0.9560 & 0.9450 & 0.9500 & 0.9330  \\\cmidrule{2-8}

& \multirow{2}{*}{SQTT} & SD   &0.7551 & 0.3226 & 0.2200 & 0.3161 & 0.7625\\   
& & Coverage & 0.9430 & 0.9590 & 0.9440 & 0.9560 & 0.9330\\   

\bottomrule

\end{tabular}
\end{table}

\subsection{Dependent Errors}

We modify the DGP \eqref{simeq} by making $u_{it}$ to be correlated to its last period value $u_{i,t-1}$ and its $2J$ neighbors, with a similar spirit in \cite{chen2021quantile}. That is:
\begin{align} 
    u_{it} = 0.2u_{i,t-1} + e_{it} + 0.2\sum_{j=i-J,j\neq i}^{i+J}e_{jt}, 
\end{align}
where $e_{it}$ $\stackrel{\text{i.i.d}}{\sim}$ $t$(3).
We arrive at similar results as before, that both \emph{NQTT} and \emph{SQTT} are consistent at all quantiles, and perform much than the \emph{GSCM} except in $\tau=0.5$ where all methods are well behaved. Moreover, inferences for \emph{NQTT} and \emph{SQTT} are valid. See Table \ref{tb3} and \ref{tb4}.

\begin{table}[!t]
\centering
\caption{Bias and RMSE: Dependent Errors}
\label{tb3}
\begin{tabular}{cccccccccc}
  \toprule
   \multicolumn{3}{c}{$\tau$} & 0.1 & 0.25 & 0.5 & 0.75 & 0.9  \\ 
   
  \midrule

\multirow{8}{*}{\makecell{N=50 \\ T=100}} 
& \multirow{2}{*}{NQTT} & Bias &0.0992 & -0.1313 & -0.0301 & 0.0605 & -0.1875   \\
& & RMSE &1.1299 & 0.6823 & 0.5442 & 0.6599 & 1.0736   \\ \cmidrule{2-8}

& \multirow{2}{*}{SQTT} & Bias &0.0503 & -0.1451 & -0.0353 & 0.0707 & -0.1410   \\
& & RMSE &1.1254 & 0.6918 & 0.5331 & 0.6657 & 1.0692 \\ \cmidrule{2-8}

& \multirow{2}{*}{Oracle} & Bias &-0.0302 & -0.0351 & -0.0319 & -0.0220 & -0.0514     \\
& & RMSE & 1.0202 & 0.6599 & 0.5641 & 0.6475 & 0.9239   \\ 
\cmidrule{2-8}

& \multirow{2}{*}{GSCM}  & Bias &-2.1746 & -0.8941 & -0.0390 & 0.8246 & 2.1245   \\
& & RMSE &2.5552 & 1.1529 & 0.5331 & 1.0764 & 2.5152   \\
\midrule

\multirow{8}{*}{\makecell{N=100 \\ T=200}} 
& \multirow{2}{*}{NQTT} & Bias &0.2857 & -0.0529 & -0.0018 & 0.0624 & -0.2728    \\
& & RMSE & 0.7905 & 0.4581 & 0.3486 & 0.4382 & 0.7930  \\ \cmidrule{2-8}

& \multirow{2}{*}{SQTT} & Bias &0.2584 & -0.0584 & -0.0023 & 0.0670 & -0.2669  \\
& & RMSE &0.8059 & 0.4606 & 0.3508 & 0.4370 & 0.8067  \\ \cmidrule{2-8}

& \multirow{2}{*}{Oracle} & Bias &0.0465 & 0.0178 & 0.0004 & 0.0009 & -0.0279     \\
& & RMSE &0.6601 & 0.4494 & 0.3719 & 0.4270 & 0.6615   \\ \cmidrule{2-8}

& \multirow{2}{*}{GSCM} & Bias &-2.2360 & -0.8974 & -0.0072 & 0.8838 & 2.2339  \\
& & RMSE & 2.4204 & 1.0248 & 0.3487 & 1.0133 & 2.4342  \\
\midrule

\multirow{8}{*}{\makecell{N=200 \\ T=400}} 

& \multirow{2}{*}{NQTT} & Bias &0.1482 & -0.0154 & 0.0060 & 0.0270 & -0.1222   \\
& & RMSE &0.5593 & 0.3090 & 0.2554 & 0.3001 & 0.5487  \\ \cmidrule{2-8}

& \multirow{2}{*}{SQTT} & Bias &0.1576 & -0.0095 & 0.0030 & 0.0179 & -0.1514   \\
& & RMSE &0.5474 & 0.3099 & 0.2545 & 0.2996 & 0.5550 \\ \cmidrule{2-8}

& \multirow{2}{*}{Oracle}  & Bias   &0.0271 & 0.0143 & 0.0010 & -0.0063 & -0.0124     \\
& & RMSE  &0.4674 & 0.3042 & 0.2653 & 0.2965 & 0.4751    \\
\cmidrule{2-8}

& \multirow{2}{*}{GSCM}  & Bias   & -2.1556 & -0.8674 & 0.0045 & 0.8814 & 2.2154  \\
& & RMSE & 2.2733 & 0.9386 & 0.2568 & 0.9504 & 2.3392  \\
\bottomrule

\end{tabular}
\end{table}

\begin{table}[!t] 
\centering
\caption{Standard Deviation and 95\% Coverage for Bootstrap: Dependent Errors}
\label{tb4}
\begin{tabular}{ccccccccc}
  \toprule
  & \multicolumn{2}{c}{$\tau$} & 0.1 & 0.25 & 0.5 & 0.75 & 0.9  \\ 
   
  \midrule

\multirow{4}{*}{\makecell{N=50 \\ T=100}}  

& \multirow{2}{*}{NQTT} & SD   & 1.2696 & 0.6756 & 0.5362 & 0.6893 & 1.2616 \\   
& & Coverage &0.9270 & 0.9450 & 0.9280 & 0.9520 & 0.9450  \\  \cmidrule{2-8}

& \multirow{2}{*}{SQTT} & SD   &1.2768 & 0.6854 & 0.5385 & 0.6885 & 1.2631 \\   
&  & Coverage &0.9260 & 0.9400 & 0.9350 & 0.9580 & 0.9430 \\   
\midrule

\multirow{4}{*}{\makecell{N=100 \\ T=200}}  

& \multirow{2}{*}{NQTT} & SD   &0.7723 & 0.4543 & 0.3606 & 0.4560 & 0.7841  \\    
& & Coverage &0.9160 & 0.9550 & 0.9390 & 0.9540 & 0.9010 \\ \cmidrule{2-8}

& \multirow{2}{*}{SQTT} & SD   &0.7650 & 0.4547 & 0.3607 & 0.4532 & 0.7778\\   
&  & Coverage &0.9180 & 0.9530 & 0.9400 & 0.9510 & 0.9050  \\   \midrule

\multirow{4}{*}{\makecell{N=200 \\ T=400}}  

& \multirow{2}{*}{NQTT} & SD   & 0.5376 & 0.3115 & 0.2514 & 0.3127 & 0.5417  \\
& & Coverage & 0.9250 & 0.9430 & 0.9330 & 0.9580 & 0.9290 \\ \cmidrule{2-8}

& \multirow{2}{*}{SQTT} & SD   &0.5397 & 0.3120 & 0.2530 & 0.3133 & 0.5480 \\   
&  & Coverage &0.9340 & 0.9470 & 0.9310 & 0.9580 & 0.9270 \\   
\bottomrule

\end{tabular}
\end{table}

\subsection{More Complicated Factor Models}
Lastly, we consider a more complicated quantile factor structure, in the same vein as \cite{ando2020quantile}. To generate the data, we first randomly draw $T \times (N+1)$ observations from the $\mathcal{U}$(0,1) distribution, denoted as $v_{it}$, and $u_{it} = \Phi^{-1}(v_{it})$, where $\Phi(\cdot)$ is the cumulative distribution function of the standard normal distribution. The following DGP of the untreated potential outcome would allow the factor structure to be quantile determined:
\begin{equation} \label{simeq_sc}
    y^0_{it} = \left\{
    \begin{aligned}
      &  \lambda_{1i}f_{1t} + \lambda_{2i}f_{2t}  + \lambda_{3i}f_{3t}  + \lambda_{6i}f_{6t} u_{it},    \hspace{4cm} \text{if $v_{it} \leq 0.3$,}       \\   
      &  \lambda_{1i}f_{1t} + \lambda_{2i}f_{2t}  + \lambda_{3i}f_{3t}  + \lambda_{4i}f_{4t} + \lambda_{6i}f_{6t} u_{it},  \hspace{2.5cm} \text{if $v_{it} \leq 0.8$,}\\ 
      &  \lambda_{1i}f_{1t} + \lambda_{2i}f_{2t}  + \lambda_{3i}f_{3t}  + \lambda_{4i}f_{4t} + \lambda_{5i}f_{5t} + \lambda_{6i}f_{6t} u_{it}, \hspace{1cm} \text{otherwise.} 
    \end{aligned}
    \right.
\end{equation}
where $f_{1t}=0.8f_{1,t-1}+\nu_{1t}$, $f_{2t}=0.5f_{2,t-1}+\nu_{2t}$, $f_{6t}=|g_{t}|$, $f_{3t}$, $f_{4t}$, $f_{5t}$, $\lambda_{1i}$, $\lambda_{2i}$, $\lambda_{3i}$, $\lambda_{4i}$, $\lambda_{5i}$, $\nu_{1t}$, $\nu_{2t}$, $g_{t}$, $u_{it}$ $\stackrel{\text{i.i.d}}{\sim}$ $\mathcal{N}$(0,1), and $\lambda_{6i}$ $\stackrel{\text{i.i.d}}{\sim}$ $\mathcal{U}$(1,2). The dimension of the common factors now varies across $i$ and $t$. Equivalently, the DGP can be written as:
\begin{equation}
    Q_{v_{it}}(y^0_{it}|f_t(\tau)) =\lambda_i'(v_{it})f_t(v_{it}),
\end{equation}
where 
\begin{equation} 
    f_t(v_{it}) = \left\{
    \begin{aligned}
      &  (f_{1t}, f_{2t}, f_{3t}, f_{6t})',    \hspace{2.6cm} \text{if $v_{it} \leq 0.3$,}       \\   
      &  (f_{1t}, f_{2t}, f_{3t},  f_{4t}, f_{6t})',  \hspace{2cm} \text{if $v_{it} \leq 0.8$,}\\ 
      &  (f_{1t}, f_{2t}, f_{3t},  f_{4t}, f_{5t}, f_{6t})', \hspace{1.4cm} \text{otherwise.} 
    \end{aligned}
    \right.
\end{equation}
\begin{equation} 
    \lambda_i(v_{it}) = \left\{
    \begin{aligned}
      &  (\lambda_{1i}, \lambda_{2i}, \lambda_{3i}, \lambda_{6i}u_{it})',    \hspace{2.6cm} \text{if $v_{it} \leq 0.3$,}       \\   
      &  (\lambda_{1i}, \lambda_{2i}, \lambda_{3i}, \lambda_{4i}, \lambda_{6i}u_{it})',  \hspace{2cm} \text{if $v_{it} \leq 0.8$,}\\ 
      &  (\lambda_{1i}, \lambda_{2i}, \lambda_{3i}, \lambda_{4i}, \lambda_{5i},  \lambda_{6i}u_{it})', \hspace{1.4cm} \text{otherwise.} 
    \end{aligned}
    \right.
\end{equation}
In this simulation, we refer \emph{Oracle} to the standard quantile regression with regressors being $(f_{1t}, f_{2t}, f_{3t},  f_{4t}, f_{5t}, f_{6t})'$. \emph{GSCM} selects the optimal number of factors from $\widehat{r}_{IC} = \argmin_{r \in \{2,\dots,8\}}  \text{IC}(r)$, since we have a larger group of factors. Table \ref{tb5} indicates that the two QTT estimators still outperform $\emph{GSCM}$ at all quantiles except $\tau=0.5$, even though the convergence towards the \emph{Oracle} is not as good as the simple setups. At $\tau=0.5$ we see that all methods are valid as before. Table \ref{tb6} verifies the validity of inference.

\begin{table}[!t]
\centering
\caption{Bias and RMSE: Quantile Variant Structures}
\label{tb5}
\begin{tabular}{cccccccccc}
  \toprule
   \multicolumn{3}{c}{$\tau$} & 0.1 & 0.25 & 0.5 & 0.75 & 0.9  \\ 
   
  \midrule

\multirow{8}{*}{\makecell{N=50 \\ T=100}} 
& \multirow{2}{*}{NQTT} & Bias &0.3691 & -0.0578 & -0.0284 & 0.1462 & -0.3522   \\
& & RMSE &0.7039 & 0.6413 & 0.3879 & 0.6646 & 0.7772   \\ \cmidrule{2-8}

& \multirow{2}{*}{SQTT} & Bias & 0.3656 & -0.0966 & -0.0194 & 0.2343 & -0.3577  \\
& & RMSE &0.7092 & 0.6730 & 0.3844 & 0.7144 & 0.7788 \\ \cmidrule{2-8}

& \multirow{2}{*}{Oracle} & Bias &0.0422 & 0.0315 & -0.0035 & -0.0696 & -0.0427     \\
& & RMSE & 0.4334 & 0.3452 & 0.3143 & 0.3696 & 0.4759   \\ 
\cmidrule{2-8}

& \multirow{2}{*}{GSCM}  & Bias &-1.2250 & -0.5291 & -0.0185 & 0.5211 & 1.3477  \\
& & RMSE &1.3681 & 0.6550 & 0.3230 & 0.6649 & 1.5005  \\
\midrule

\multirow{8}{*}{\makecell{N=100 \\ T=200}} 
& \multirow{2}{*}{NQTT} & Bias &0.2190 & 0.0473 & -0.0515 & 0.0169 & -0.2241    \\
& & RMSE &0.5070 & 0.3108 & 0.2372 & 0.3553 & 0.5708   \\ \cmidrule{2-8}

& \multirow{2}{*}{SQTT} & Bias & 0.2437 & 0.0096 & -0.0519 & 0.0637 & -0.2380  \\
& & RMSE &0.5186 & 0.3210 & 0.2363 & 0.3731 & 0.5889 \\ \cmidrule{2-8}

& \multirow{2}{*}{Oracle} & Bias &0.0160 & 0.0150 & -0.0106 & -0.0602 & -0.0064    \\
& & RMSE & 0.3611 & 0.2932 & 0.2521 & 0.2737 & 0.3767 \\ \cmidrule{2-8}

& \multirow{2}{*}{GSCM} & Bias & -1.7605 & -0.7512 & -0.0351 & 0.7084 & 1.8583  \\
& & RMSE & 1.8339 & 0.8175 & 0.2431 & 0.7861 & 1.9428   \\
\midrule

\multirow{8}{*}{\makecell{N=200 \\ T=400}} 

& \multirow{2}{*}{NQTT} & Bias &0.1535 & 0.0508 & -0.1106 & 0.0036 & -0.1846   \\
& & RMSE &0.3081 & 0.1976 & 0.1826 & 0.2973 & 0.3470  \\ \cmidrule{2-8}

& \multirow{2}{*}{SQTT} & Bias &0.1530 & 0.0278 & -0.1106 & 0.0440 & -0.1801  \\
& & RMSE &0.3146 & 0.1967 & 0.1824 & 0.3028 & 0.3538 \\ \cmidrule{2-8}

& \multirow{2}{*}{Oracle}  & Bias   &0.0228 & 0.0071 & -0.0401 & 0.0138 & -0.0268    \\
& & RMSE  & 0.2379 & 0.1823 & 0.1615 & 0.1758 & 0.2284  \\
\cmidrule{2-8}

& \multirow{2}{*}{GSCM}  & Bias   &-1.5012 & -0.6577 & -0.0797 & 0.6720 & 1.5903  \\
& & RMSE & 1.5385 & 0.6894 & 0.1729 & 0.7124 & 1.6231   \\
\bottomrule

\end{tabular}
\end{table}

\begin{table}[!t] 
\centering
\caption{Standard Deviation and 95\% Coverage for Bootstrap: Quantile Variant Structures}
\label{tb6}
\begin{tabular}{ccccccccc}
  \toprule
  & \multicolumn{2}{c}{$\tau$} & 0.1 & 0.25 & 0.5 & 0.75 & 0.9  \\ 
   
  \midrule

\multirow{4}{*}{\makecell{N=50 \\ T=100}}  

& \multirow{2}{*}{NQTT} & SD   &0.6723 & 0.5246 & 0.3960 & 0.5377 & 0.7517   \\   
& & Coverage &0.9250 & 0.8930 & 0.9510 & 0.8810 & 0.9340  \\  \cmidrule{2-8}

& \multirow{2}{*}{SQTT} & SD   &0.6827 & 0.5229 & 0.3957 & 0.5363 & 0.7619\\   
&  & Coverage &0.9270 & 0.8770 & 0.9590 & 0.8550 & 0.9350 \\   
\midrule

\multirow{4}{*}{\makecell{N=100 \\ T=200}}  

& \multirow{2}{*}{NQTT}& SD   &0.4717 & 0.3239 & 0.2452 & 0.3663 & 0.5598  \\    
& & Coverage &0.9370 & 0.9570 & 0.9400 & 0.9550 & 0.9180 \\ \cmidrule{2-8}

& \multirow{2}{*}{SQTT} & SD   &0.4807 & 0.3312 & 0.2433 & 0.3665 & 0.5643\\   
&  & Coverage &0.9190 & 0.9470 & 0.9410 & 0.9410 & 0.9380 \\  \midrule

\multirow{4}{*}{\makecell{N=200 \\ T=400}}  

& \multirow{2}{*}{NQTT}& SD   &0.2795 & 0.1952 & 0.1442 & 0.2762 & 0.3151   \\
& & Coverage &0.9210 & 0.9530 & 0.8800 & 0.9440 & 0.9340  \\ \cmidrule{2-8}

& \multirow{2}{*}{SQTT} & SD   &0.2897 & 0.1988 & 0.1444 & 0.2818 & 0.3264\\   
&  & Coverage &0.9250 & 0.9510 & 0.8810 & 0.9520 & 0.9280 \\  
\bottomrule

\end{tabular}
\end{table}

\section{Empirical Example: California’s
Tobacco Control} \label{sec:cali}
In this section, we revisit the well-known case study of California’s tobacco control program, originally analyzed by \citet{abadie2010synthetic}, and re-examine it using our proposed method.

Proposition 99, enacted in 1988, marked the launch of the first large-scale modern tobacco control initiative in the United States. It increased the cigarette excise tax by 25 cents per pack and allocated substantial funding toward anti-smoking campaigns, public health efforts, and local policy enforcement. This intervention catalyzed broad legislative and behavioral changes in California’s tobacco environment.

We use annual state-level panel data from 1970 to 2000\footnote{The data comes from the R package `synthdid' of \cite{arkhangelsky2021synthetic}.}, with per capita cigarette sales (measured in packs) as the outcome variable, and ignore other covariates. The control group includes 38 states, while others with similar anti-tobacco programs or large tax hikes are excluded. California is the treated unit. We have $T_0=19$ pre-treatment years, and $T_1=12$ post-treatment years.

\begin{figure}[!t]
\centering
\caption{QTT: California Proposition 99}
\includegraphics[scale=0.6]{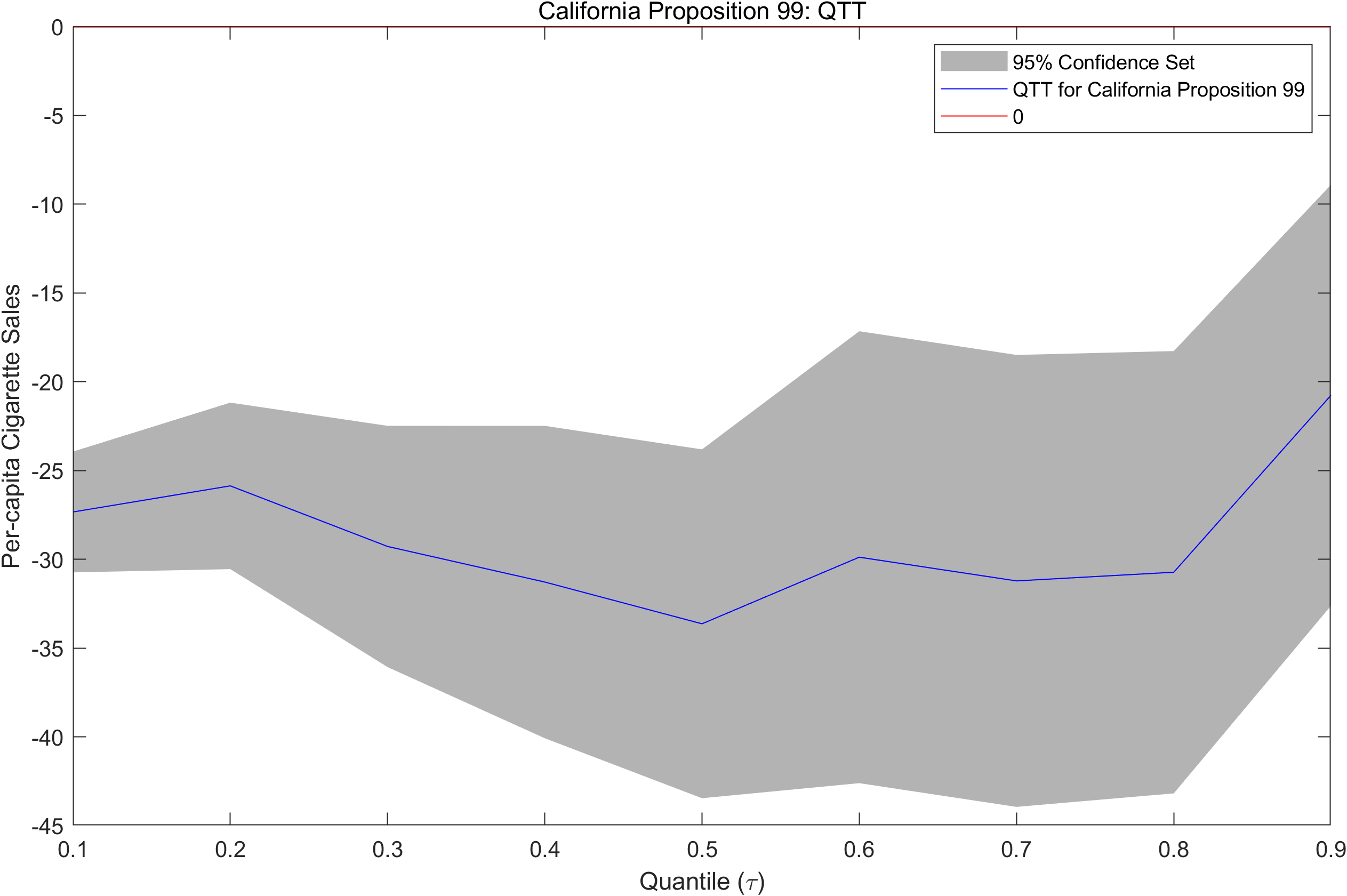}
\caption*{Note: The blue line is the estimated QTT, and the shaded area represents the $95\%$ confidence interval.}
\label{cali_QTT}
\end{figure}

Figure \ref{cali_QTT} further explores the impact of California’s Proposition 99 at deciles ranging from the 10th to the 90th percentile using NQTT. These estimates suggest that the tobacco control intervention had a consistently negative effect across the entire distribution of smoking behavior, with magnitudes ranging from a reduction of approximately 20.76 packs at the 90th percentile to 33.63 packs at the median. Compared to the findings of \citet{abadie2010synthetic}, who report that cigarette consumption declined by an average of nearly 20 packs per capita over 1989–2000, our results indicate a stronger overall effect of Proposition 99 on cigarette consumption. In particular, the QTTs suggest that the policy had a broad-based impact across the entire distribution, affecting both peak and trough years of cigarette consumption within the post-treatment period.

Moreover, we note that the confidence intervals obtained via bootstrap methods are relatively wide compared to those reported in the empirical example in the main text. This likely reflects the limited number of time periods available for estimation.

\end{document}